\documentclass[fleqn,usenatbib]{mnras}

\usepackage{newtxtext,newtxmath}

\usepackage[T1]{fontenc}

\DeclareRobustCommand{\VAN}[3]{#2}
\let\VANthebibliography\thebibliography
\def\thebibliography{\DeclareRobustCommand{\VAN}[3]{##3}\VANthebibliography}

\usepackage{graphicx}	
\usepackage{amsmath}	
\usepackage{xspace}
\usepackage[dvipsnames,x11names]{xcolor}

\newcommand{\astraeus}{\textsc{Astraeus}\xspace}
\newcommand{\delphi}{\textsc{Delphi}\xspace}
\newcommand{\cifog}{\textsc{Cifog}\xspace}
\newcommand{\vsmdpl}{\textsc{VSMDPL}\xspace}

\newcommand{\Msun}{\ifmmode \mbox{M}_{\sun} \else $\mbox{M}_{\sun}$\xspace\fi}
\newcommand{\Zsun}{\ifmmode \mbox{Z}_{\sun} \else $\mbox{Z}_{\sun}$\xspace\fi}
\newcommand{\Mbh}{\ifmmode M_{\bullet} \else $M_\bullet$\xspace\fi}
\newcommand{\Mstar}{\ifmmode M_{\star} \else $M_\star$\xspace\fi}
\newcommand{\MUV}{\ifmmode M_{\rm UV} \else $M_{\rm UV}$\xspace\fi}
\newcommand{\fesc}{\ifmmode f_{\rm esc} \else $f_{\rm esc}$\xspace\fi}
\newcommand{\fescAGN}{\ifmmode f_{\rm esc}^{\rm AGN} \else $f_{\rm esc}^{\rm AGN}$\xspace\fi}
\newcommand{\fescstar}{\ifmmode f_{\rm esc}^{\star} \else $f_{\rm esc}^{\star}$\xspace\fi}
\newcommand{\fedd}{\ifmmode f_{\rm Edd} \else $f_{\rm Edd}$\xspace\fi}

\defcitealias{Hutter2021}{Paper~I}
\defcitealias{Ucci2021b}{Paper~V}

\title[AGN assembly and reionization]{Astraeus VI: Hierarchical assembly of AGN and their large-scale effect during the Epoch of Reionization}

\author[M. Trebitsch et al.]{
  Maxime Trebitsch,$^{1}$\thanks{E-mail: m.trebitsch@rug.nl}
  Anne Hutter$^{1}$,
  Pratika Dayal$^{1}$,
  Stefan Gottlöber$^{2}$,
  Laurent Legrand$^{1}$ and
  Gustavo Yepes$^{3,4}$
\\
$^{1}$Kapteyn Astronomical Institute, University of Groningen, P.O. Box 800, 9700 AV Groningen, The Netherlands\\
$^{2}$Leibniz-Institut für Astrophysik, An der Sternwarte 16, 14482 Potsdam, Germany\\
$^{3}$Departamento de Fısica Teorica, Modulo 8, Facultad de Ciencias, Universidad Autonoma de Madrid, 28049 Madrid, Spain\\
$^{4}$CIAFF, Facultad de Ciencias, Universidad Autonoma de Madrid, 28049 Madrid, Spain
}

\date{Accepted XXX. Received YYY; in original form ZZZ}

\pubyear{2021}

\begin{document}
\label{firstpage}
\pagerange{\pageref{firstpage}--\pageref{lastpage}}
\maketitle

\begin{abstract}
  In this work, the sixth of a series, we use the \astraeus (semi-numerical rAdiative tranSfer coupling of galaxy formaTion and Reionization in N-body dark matter simUlationS) framework to investigate the nature of the sources that reionized the Universe.
  We extend \astraeus, which already couples a galaxy formation semi-analytical model with a detailed semi-numerical reionization scheme, to include a model for black hole formation, growth, and the production of ionizing radiation from associated AGN (active galactic nuclei).
  We calibrate our fiducial AGN model to reproduce the bolometric luminosity function at $z \simeq 5$, and explore the role of the resulting AGN population in reionizing the Universe.
  We find that in all the models yielding a reasonable AGN luminosity function, galaxies dominate overwhelmingly the ionizing budget during the Epoch of Reionization, with AGN accounting for 1-10\% of the ionizing budget at $z=6$ and starting to play a role only below $z \lesssim 5$.
\end{abstract}

\begin{keywords}
galaxies: high-redshift -- dark ages, reionization, first stars -- intergalactic medium -- galaxies: active -- methods: numerical
\end{keywords}


\section{Introduction}

During its first billion years, the Universe is the stage of major transformations for its baryonic content. The first stars and black holes form at $z \lesssim 30$, and the intense ultraviolet (UV) radiation they produce gradually ionizes the hydrogen in the intergalactic medium (IGM), creating ionized bubbles that grow for about 1 Gyr, until they fully overlap at $z \simeq 6$ \citep[e.g.][]{Fan2006b}: this is the Epoch of Reionization. Current observational constraints suggest a late and relatively rapid reionization process \citep[e.g.][]{Planck2018}, with its tail end extending below $z\lesssim 6$ \citep[e.g.][]{Kashino2020,Bosman2022}.
Amongst the different sources that have been proposed to contribute to the photon budget of reionization, two of them have particularly stood out: young, massive stars in galaxies and active galactic nuclei (AGN) powered by the accretion onto super-massive black holes (SMBHs).
The census of the sources responsible for producing the bulk of the ionizing photons that are responsible for reionizing the Universe has been the focus of significant observational and theoretical work \citep[see e.g.][]{Dayal2018}.

Current models suggest that the sheer number of galaxies make them the main drivers of reionization \citep[e.g.][]{Becker2013,Robertson2015,Madau2017a,Dayal2020}, but understanding which galaxies are the main contributors is still an open question. While a significant contribution from faint galaxies seems to be required \citep[e.g.][]{Duncan2015,Robertson2015,Hutter2021b}, and especially to explain the end of reionization \citep[e.g.][]{Kakiichi2018,Meyer2019,Ocvirk2021}, this might lead to a too slow reionization \citep[e.g.][]{Finkelstein2019}, and several studies have hinted at a significant contribution of slightly brighter, more common, $\MUV \simeq -19$ galaxies \citep[e.g.][]{Naidu2020,Naidu2021,Matthee2022}.
Adding to the complexity, the role of AGN has recently been revisited by multiple studies focusing on the faint-end of the AGN luminosity function (LF) at $z \simeq 4-6$. The observations of \citet{Giallongo2015,Giallongo2019} and \citet{Boutsia2018} have hinted at a larger than expected number density of faint AGN at $z \gtrsim 4$, which could imply a significant contribution of AGN to the establishment of the ionizing UV background if such number densities hold up to higher redshifts \citep[e.g.][]{Grazian2018,Mitra2018}. This scenario has been heavily debated in the past few years, with other studies finding lower AGN number densities \citep[e.g.][]{Weigel2015,McGreer2018,Parsa2018,Akiyama2018}.

Nevertheless, theoretical models of reionization need to take the contribution from AGN into account.
Earlier models \citep[e.g.][]{Volonteri2009} had suggested an important contribution of AGN, but the contribution of these high-redshift AGN is very sensitive to the growth history of SMBHs. In particular, numerical simulations indicate that in low-mass galaxies, the growth of SMBHs is stunted by supernova feedback \citep[e.g.][]{Dubois2015,Habouzit2017,Prieto2017,Trebitsch2018}, which strongly limits the contribution of these AGN to the UV background. These results are in line with the findings of \citet{Dayal2020}, who used the \delphi \citep{Dayal2014} semi-analytical model coupled with a `one-zone' reionization equation and showed that AGN were sub-dominant contributors to the UV background during the Epoch of Reionization. AGN-assisted models can however have an impact on their local reionization history: rare bright sources have been suggested by e.g. \citet{Chardin2015,Chardin2017} to produce variations in the UV background that could explain the fluctuations in the Lyman-$\alpha$ effective optical depth observed at the end of reionization \citep{Becker2015}. Similarly, bright sources are expected to leave an imprint on the thermal history of the IGM \citep[e.g.][]{Eide2020}.

From a numerical standpoint, it is extremely challenging to bring together detailed galaxy formation and SMBH growth hydrodynamical simulations and large scale reionization models to assess self-consistently the contribution of AGN to the UV background and their impact on the topology of reionization. A first attempt has been made by \citet{Trebitsch2021}, who used a dedicated radiation hydrodynamics cosmological simulation and found that even in environment that are favourable for SMBH growth, the global contribution of these high-redshift AGN to reionization is subdominant. However, because they focus on a relatively small region of the Universe, they cannot assess directly the impact of AGN on the larger scales of reionization.
In this work, we take the complementary approach of modelling the galaxy and AGN population in a large volume using a physically-motivated semi-analytical model that we apply to a cosmological N-body simulation to quantify not only the amount of ionizing photons coming from AGN, but also how they are spatially distributed to fully model their impact on the large-scale reionization process.

We first describe our model in Sect.~\ref{sec:methods}, presenting in particular our new AGN implementation in Sect.~\ref{sec:methods:agn}. We then calibrate our model to reproduce the AGN bolometric LF at high-redshift and investigate the properties of the resulting AGN population in Sect.~\ref{sec:agn}. Finally, we use in Sect.~\ref{sec:results} the outcome of our reionization model to establish the role of AGN in the reionization of the Universe.

\section{Simulations and AGN model}
\label{sec:methods}

In this paper, we jointly model the formation and evolution of star-forming galaxies and AGN self-consistently coupled with reionization using the \astraeus framework \citep[][hereafter \citetalias{Hutter2021}]{Hutter2021}. This framework relies on an N-body dark matter (DM) simulation to provide a halo catalogue and merger tree, and applies an enhanced version of the \delphi \citep{Dayal2014} semi-analytical model to follow the physics of baryons, while the radiation and ionization fields are evolved with the \cifog \citep{Hutter2018} semi-numerical reionization scheme.
We start this section by presenting the N-body DM simulation that serves as a basis for this work. Then, as the \astraeus framework has been extensively described in \citetalias{Hutter2021}, we only briefly summarise in Sect.~\ref{sec:methods:astraeus} the main features of the code, and refer the interested reader to that paper for more details. Finally, our new AGN implementation is described in Sect.~\ref{sec:methods:agn}.

\subsection{N-body simulation and haloes}
\label{sec:methods:n-body}
We run our semi-analytical model on the \vsmdpl (Very Small Multi\-Dark PLanck) N-body simulation, which is part of the \textsc{MultiDark} simulation project\footnote{\label{fn:multidark}\url{https://www.cosmosim.org/}} \citep{Klypin2016}. The simulation has been run using the \textsc{Gadget-2} \citep{Springel2005} TreePM N-body code, and assumes a cosmology consistent with the \emph{Planck} 2018 results \citep{Planck2018}: $h=0.6777$, $\Omega_m=0.307115$, $\Omega_b=0.048206$, $\Omega_\Lambda=0.692885$, $n_s = 0.96$ and $\sigma_8=0.8228$.
The \vsmdpl box has a side length of $160 h^{-1}\,\mbox{Mpc}$, and follows the evolution of $3840^3$ DM particles, yielding a mass resolution of $m_{\mathrm{DM}} = 6.2\times 10^{6} h^{-1}\,\Msun$. The simulation used a fixed gravitational softening length of $2h^{-1}\,\mbox{kpc}$ (comoving) at $z>1$.
The database comprises 150 snapshots available between $z=25$ and $z=0$, and we select the first 74 of them (down to $z = 4.5$).

Haloes and subhaloes have been identified using the \textsc{Rockstar} phase-space halo finder \citep{Behroozi2013} for all 150 snapshots, requiring structures to be resolved by at least 20 particles (corresponding to a minimum halo mass of $M_{\mathrm{vir,min}} = 1.24\times 10^{8}h^{-1}\,\Msun$. From these halo catalogues, mergers trees have been produced using \textsc{Consistent Trees} \citep{Behroozi2013b}, and then resorted from a tree-branch-by-tree-branch (``vertical'') order to a redshift-by-redshift order within each tree, as described in \citetalias{Hutter2021}. In total, the final catalogue contains more than 73 million galaxies at $z = 4.5$.

Finally, the density field has been produced for all snapshots by projecting the particles onto a $2048^3$ grid, that we then have resampled to a $256^3$ grid to serve as input to the reionization module.

\subsection{The \astraeus framework}
\label{sec:methods:astraeus}
The \astraeus framework models all the key processes related to the assembly of galaxies in the high-redshift Universe: accretion of gas and DM, growth via mergers bringing in gas, DM and stars, star formation and the resulting type II supernova (SN) feedback, as well as the impact of the inhomogeneous ionizing background generated by the distribution of galaxies (and, as we introduce in this work, AGN). At each step of the simulation, the baryonic processes are coupled to the growth of the DM haloes (via merger and accretion) directly derived from the halo properties evolved in the N-body simulation.

When haloes are initialised in the simulations (which we will refer to as `starting haloes'), we assume that their initial gas content is purely set by the cosmological baryon fraction $f_b = \Omega_b/\Omega_m$. From that point, the halo growth has a merger component $M_{\rm h}^{\rm mer}$ (the sum of the masses of the resolved progenitors) and a smooth accretion component $M_{\rm h}^{\rm acc}$ (the rest), which both contribute gas at different rates: accretion is assumed to always bring a gas mas of $M_{\rm g}^{\rm acc} = f_b M_{\rm h}^{\rm acc}$, while mergers bring the amount of gas left in each progenitor after star formation, BH growth, and the corresponding feedback. For starting haloes, the merger term is simply set to zero. For haloes living in ionized regions, reionization feedback can reduce the amount of gas that can be sustained in the halo to a gas fraction $f_{\rm g}$, as discussed in \citetalias{Hutter2021}. In this case, the initial gas mass $M_{\rm g}^i(z)$ available in a halo of mass $M_{\rm h} = M_{\rm h}^{\rm mer} + M_{\rm h}^{\rm acc}$ at the beginning of a timestep at $z$ is given by
\begin{equation}
  \label{eq:Mgas_initial}
  M_{\rm g}^i(z) = \min\left[ M_{\rm g}^{\rm mer}(z) + M_{\rm g}^{\rm acc}(z), f_{\rm g} \frac{\Omega_b}{\Omega_m} M_{\rm h}(z) \right].
\end{equation}
Following \citet[][hereafter \citetalias{Ucci2021b}]{Ucci2021b}, we follow the metallicity of the gas in the haloes in addition to the IGM metallicity. The accretion component is assumed to proceed at the self-consistently evolved average IGM metallicity, and each progenitor brings its own metal component.

In a halo, star formation and SN feedback are coupled together such that the star formation proceeds at an effective efficiency $f_{\star}^{\rm eff}$ set by the feedback strength.
The amount of newly formed stars is given by $\Mstar^{\rm new}(z) = f_\star^{\rm eff} M_{\rm g}^i(z)$. 
The evolution of this newly formed stellar population results in SN that are all assumed to release an energy of $E_{51} = 10^{51}\,\mbox{erg}$, and we couple a fraction $f_w = 0.2$ of that energy to the gas reservoir. As described in \citetalias{Ucci2021b}, we use a `delayed feedback' scheme that accounts for the mass-dependent lifetimes of stars. To avoid introducing an artificial time sampling of the star formation histories, we need to further assume that star formation is continuous and uniformly distributed over each timestep of the simulation. At any time the intrinsic star formation efficiency $f_\star$ is normalised so that the amount of stars formed over a timestep correspond to an efficiency of $f_\star^0 = 0.025$ over $20\,\mbox{Myr}$.
As in \citet{Dayal2014} and \citetalias{Hutter2021}, we cap the star formation efficiency to the minimum efficiency required to eject all the gas left in the halo after star formation:
\begin{equation}
  \label{eq:ejected_fraction}
  f_\star^{\rm ej} = \frac{\Mstar^{\rm new}(z)}{\Mstar^{\rm new}(z) + M_{\rm g}^{\rm ej}(z)}.
\end{equation}
For instantaneous feedback, this would reduce to $f_\star^{\rm ej} = v_c^2 / (v_c^2 + f_w E_{51} \nu)$, where $v_c$ is the circular velocity of the halo and $\nu = 0.0077\,\Msun^{-1}$ is the number of SN per stellar mass for the \citet{Salpeter1955} IMF between 0.1 and 100 \Msun that we assume in this work.
Finally, we set the effective star-formation efficiency to $f_\star^{\rm eff} = \min\left[f_\star^0, f_\star^{\rm ej}\right]$.
As a result, the star-formation efficiency in low-mass galaxies is capped at the efficiency required to eject all of the remaining gas, while it still saturates to the threshold value at higher masses \citep[see e.g. Fig.~1 of][]{Legrand2022}.
After star formation and feedback, the amount of gas left in the halo is given by
\begin{equation}
  \label{eq:Mgas_star}
  M_{\rm g}^\star = \left(M_{\rm g}^i - M_\star^{\rm new}\right)\left(1 - \frac{f_\star^{\rm eff}}{f_\star^{\rm ej}}\right).
\end{equation}

We model the radiation output from each galaxy in order to evolve self-consistently the inhomogeneous ionizing background using \cifog. As in \citetalias{Hutter2021}, we assign each galaxy a spectrum by convolving its star formation history with a starburst spectrum obtained with the \textsc{Starburst99} \citep{Leitherer1999} stellar population synthesis model (for the impact of this model, see \citetalias{Hutter2021}). For simplicity, we assume a low $Z = 0.05\,\Zsun$ metallicity for all galaxies when evaluating their ionizing output\footnote{\label{fn:SED-Z}Even by $z \simeq 5$, our most massive galaxies have a metallicity of at most $0.25\,\Zsun$ \citepalias{Ucci2021b}. Since the ionizing production rate only changes by a factor $\simeq 1.1$ between metallicity values of $0.05$ and $0.25\,\Zsun$, we do not expect our results to be affected by this choice.}. The intrinsic ionizing emissivity of each galaxy is then given by the integral of its spectral energy distribution (SED) in the \textsc{Hi}-ionizing band ($\lambda < 912\,\mbox{Å}$). Only a fraction \fescstar of these photons will actually make it to the IGM and contribute to reionization. Motivated by the simulations that find a strong connection between SN feedback and escape of ionizing radiation \citep[e.g.][]{Wise2014, Kimm2014, Trebitsch2017}, we assume that \fescstar scales with the strength of SN feedback in our model:
\begin{equation}
  \label{eq:fescstar}
  \fescstar = {\fescstar}^0 \frac{f_\star^{\rm eff}}{f_\star^{\rm ej}},
\end{equation}
where ${\fescstar}^0 = 0.24$ is a normalisation chosen to reproduce the CMB constraints on the Thomson optical depth of reionization. This model leads naturally to a larger contribution of low-mass galaxies (for which $f_\star^{\rm eff} \simeq f_\star^{\rm ej}$) to reionization, as also shown in \citet{Hutter2021b}.

The resulting ionizing background will ionize and heat the IGM, both reducing the amount of gas available for star formation \citep[e.g.]{Barkana1999, Shapiro2004} and increasing the Jeans mass. The latter effect increases the minimum mass for galaxy formation, thus reducing the amount of accreted gas onto the galaxy \citep[e.g.][]{Couchman1986, Efstathiou1992}. We regroup both these phenomena under the `radiative feedback' umbrella term, and follow the `photo-ionization model' of \citetalias{Hutter2021}, based on the estimate of \citet{Sobacchi2013} for the critical mass below which the gas fraction $f_{\rm g}$ is suppressed. In all the reionization feedback model explored in \citetalias{Hutter2021}, this is the most intermediate one.

We calibrate our galaxy formation parameters ($f_\star^0, f_w$) to reproduce the galaxy UV LF in the high-redshift Universe, as well as the stellar mass function and the derived SFR and stellar mass densities. Doing so, we need to attenuate the simulated galaxy UV LF: we use for this the dust model that has been implemented in \astraeus and coupled to the metal evolution model of \citetalias{Ucci2021b}. This dust model is extremely similar to the one implemented in \delphi (see Dayal et al, submitted), and yields a dust mass $M_{\rm d}$ in each galaxy. For each galaxy, we compute the intrinsic UV luminosity $L_{\rm UV}^{\rm int}$ around $1500\,\mbox{Å}$ in the same way that we compute its ionizing emissivity. We then assume dust, stars, and gas are co-spatial and homogeneously distributed in a disc of radius $r_{\rm g} = 4.5 \lambda R_{\rm vir}$ \citep{Ferrara2000}, where $\lambda = 0.04$ \citep[e.g.][]{Bullock2001} is the spin parameter of the halo and $R_{\rm vir}$ is the virial radius of the halo at the redshift of interest. The dust optical depth is then given by
\begin{equation}
  \label{eq:tau_dust}
  \tau_{\rm d} = \frac{3 M_{\rm d}}{4\pi r_{\rm g}^2 a s},
\end{equation}
with $a = 0.05\,\mu\mbox{m}$ the grain size and $s = 2.25\,\mbox{g}\,\mbox{cm}^{-3}$ the density, appropriate for carbonaceous grains.
Considering the disc as a slab, we can compute the escape fraction of (non-ionizing) UV photons as
\begin{equation}
  \label{eq:uv_escape}
  f_{\rm d} = \frac{1-e^{-\tau_{\rm d}}}{\tau_{\rm d}},
\end{equation}
and the observed galaxy luminosity is therefore $L_{\rm UV} = f_{\rm d} L_{\rm UV}^{\rm int}$. Calibrating our model on the UV LF from \citet{Bouwens2017, Livermore2017, Atek2018}, we find that ($f_\star^0 = 0.025, f_w = 0.2$) gives the best results.

\subsection{AGN model}
\label{sec:methods:agn}

We improve on the previous \astraeus implementation by including a physically motivated model for AGN that describes the seeding of SMBH in high-$z$ haloes, their growth through mergers and accretion, and the resulting feedback. A similar model has been implemented in \delphi \citep{Dayal2019, Piana2021}, albeit with a simpler seeding prescription. The main improvement compared to these work is that within the \astraeus framework, the ionizing radiation produced by AGN is self-consistently coupled to the spatially varying ionization field, allowing us to investigate directly the role of AGN in reionizing the Universe.

\begin{table}
  \centering
  \caption{Model parameters and chosen values in this work. The parameters that we focus on are marked in bold.}
  \label{tab:example_table}
  \begin{tabular*}{\columnwidth}{ccc}
    \hline\hline
    Parameter & Value or reference & Description\\
    \hline
    $f_\star^0$ & $0.025$ & Maximum star-formation efficiency \\
    $f_w$      & $0.2$   & SN coupling efficiency \\
    ${\fescstar}^0$ & 0.24 & Galaxy escape fraction \\
    - & Photo-ionization & Radiative feedback model \\
    IMF & \citet{Salpeter1955} & For stellar evolution, enrichment, SED \\
    SED & \textsc{Starburst99} & ionizing SED model \\
    \hline
    $Z_{\rm crit}$ & $1.58\times 10^{-4}\,\Zsun$ & Critical metallicity for BH seeding\\
    $\mathcal{D}_{\rm crit}$ & $4.4\times 10^{-9}$ & Critical dust for BH seeding\\
    $J_{\rm crit}$ & $30 - 300\, J_{21}$ & Critical $J_{\rm LW}$ for DCBH seeding\\
    $M_{\rm PopIII}$ & $150h^{-1}\,\Msun$ & Pop III seed mass\\
    $M_{\rm DCBH}$ & $10^{4-5}h^{-1}\,\Msun$ & DCBH seed mass\\
    $\Delta x_{\rm LW}$ & $39 h^{-1}\,\mbox{kpc}$ & Cell size for the LW background\\
    $f_{\rm seed}$ & $0.1, 0.2, 1.0$ & {\bf Pre-seeded self-enriched haloes}\\
    $M_{\rm crit}$ & \citet{Bower2017} & Critical halo mass for $\fedd$\\
    $\fedd^{\rm high}$ & $0.7 - 1$ & {\bf \fedd in high-mass haloes} \\
    $\fedd^{\rm low}$ & $7.5\times 10^{-4}$ & \fedd in low-mass haloes \\
    $f_{\bullet}^{\rm acc}$ & $5.5 \times 10^{-4}$& Gas fraction for BH growth\\
    $f_{\bullet}^w$ & $0.003$ & AGN coupling efficiency \\
    \fescAGN & Sect.~\ref{sec:methods:agn:radiation} & {\bf AGN escape fraction} \\
    \hline
  \end{tabular*}
\end{table}

\subsubsection{SMBH seeding}
\label{sec:methods:agn:seeding}

At each timestep, we select SMBH formation sites amongst the starting haloes. We model the formation of two types of SMBH seeds: direct-collapse black holes (DCBHs), as well as black holes remnants of massive Population III stars (hereafter Pop III BHs). For both types of seeds, models \citep[see the review of][]{Volonteri2010} require them to form from metal-free or extremely metal-poor gas with a metallicity $Z \leq Z_{\rm crit} = 1.58\times 10^{-4}\,\Zsun$ and a dust to gas ratio $\mathcal{D} \leq \mathcal{D}_{\rm crit} = 4.4\times 10^{-9}$ \citep{Omukai2000,Schneider2012}.

\begin{figure}
  \centering
  \includegraphics[width=\columnwidth]{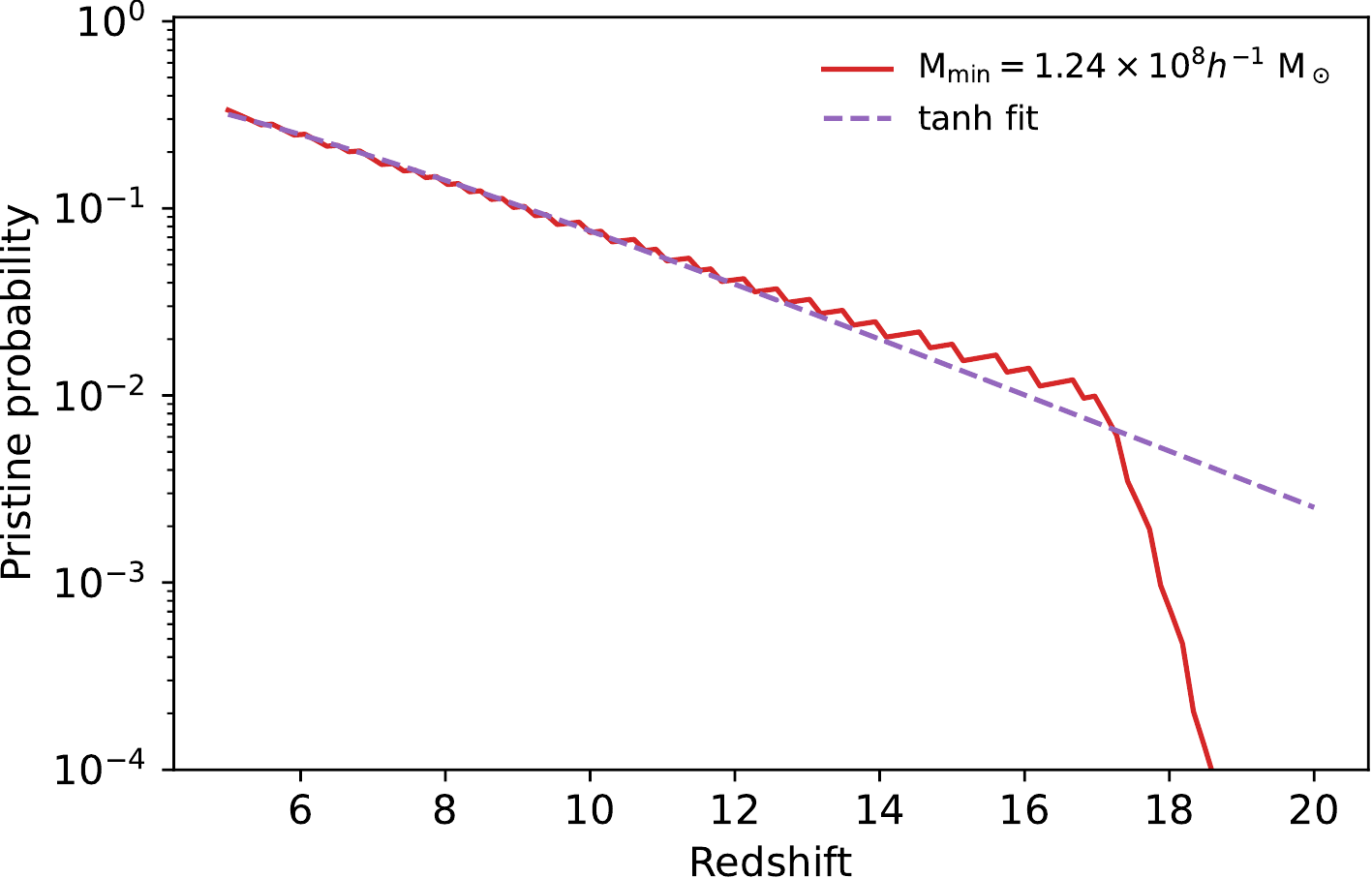}\vspace{-1em}
  \caption{Probability for a starting halo in \astraeus to not be self-enriched (solid red line) and the analytical fit (dashed purple line). In the $z\gtrsim 10$ Universe, $90-99\%$ of our starting haloes are self-enriched.}
  \label{fig:pristine_probability}
\end{figure}

Because of the resolution we employ in this work ($M_{\mathrm{vir,min}} = 1.24\times 10^{8}h^{-1}\,\Msun$), we only barely resolve atomic cooling haloes, and therefore cannot account for Pop III star formation in the minihalo progenitors of our starting haloes. We follow the method described by \citet[][see also \citealp{Dijkstra2014}]{Trenti2007,Trenti2009} to estimate the fraction of starting haloes that have been self-enriched by previous episodes of Pop III star formation. Using linear theory, they compute the probability for a halo of mass $M$ at a redshift $z$ to have had at least one progenitor massive enough to sustain $\mbox{H}_2$ or atomic cooling and early enough for Pop III stars to have formed and exploded as supernovae before $z$. Because this requires knowledge of the Lyman-Werner (LW) background contributed by stars that we do not follow, we use their estimate of the evolution of the LW flux \citep[eq 19 of][]{Trenti2009}. Using our minimum halo mass $M_{\mathrm{vir,min}}$ , we can then infer the probability of a starting halo to be self-enriched by previous star formation episodes, which we show as the red curve in Fig.~\ref{fig:pristine_probability}. For computational efficiency, we fit the resulting probability using a $\tanh$ function, which we found to give a good fit to the results:
\begin{equation}
  \label{eq:pristine_probability}
  P_{\rm prisitine} = \frac{1}{2}\left( 1 + \tanh\left(\frac{z_0-z}{\Delta z}\right)\right),
\end{equation}
with $z_0 = 2.81$ and $\Delta z = 5.75$. While it may seem counter-intuitive that the pristine probability is higher at lower redshift, it can be understood as the fact that a halo of mass $M_{\mathrm{vir,min}}$ corresponds to a higher overdensity at high redshift, and is therefore more likely to have already formed stars. We note here that this pristine probability is just the probability of not having been self-enriched, and does not account for the ``environmental'' enrichment coming from the naturally increasing metallicity of the IGM.

For pristine haloes, we then must decide whether they are hosting a DCBH seed or a Pop III seed. Models have suggested that DCBH formation is controlled by the intensity of the LW flux that must be high enough to dissociate $\mbox{H}_2$ molecules and therefore prevent fragmentation of the gas feeding the super-massive star that will end up as a DCBH \citep[][although see \citealp{Spaans2006,Lodato2006,Begelman2006,Begelman2009} for alternative models]{Bromm2003,Dijkstra2008}. The exact value of this critical flux is highly debated in the literature \citep[see e.g. the discussion in][]{Inayoshi2020}, but values range between $J_{\rm crit} = 10 - 1000\, J_{21}$ with $J_{21} = 10^{-21}\,\mbox{erg} \mbox{s}^{-1} \mbox{Hz}^{-1} \mbox{sr}^{-1} \mbox{cm}^{-2}$.
We then estimate the local LW background directly from the simulation: each galaxy in the volume is assigned a luminosity in the LW band based on its mass, age and metallicity. For galaxies with stellar population more metal-rich than $1.34\times 10^{-7} = 10^{-5}\,\Zsun$, we use the \textsc{Bpass} v2.2.1 stellar population synthesis model \citep[][]{Eldridge2017}, while we assume a Pop III spectral energy distribution from \citet{Schaerer2003} below that threshold. We checked that our choice of \textsc{Bpass} made very little difference on the resulting LW background.
At each snapshot, we estimate the local LW flux $J_{\rm LW}$ hitting each halo using the approach of \citet{Barnes1986} to estimate the gravitational field: since the IGM is optically thin to LW photons, the LW flux around each source decays as $1/r^2$ just like the gravitational force, the method is effectively the same. In practice, we choose an opening angle $\theta = 1$, and we stop the tree at level $\ell = 12$, equivalent to a cell size of $\Delta x = 39h^{-1}\,\mbox{kpc}$ (effectively grouping together particles closer than that). We tested that cutting the tree at any level from $\ell = 12$ to $\ell = 16$ made no difference on our results.
The main difference with the gravitational tree from the Barnes \& Hut method is that we further attenuated the luminosity using the picket-fence modulation factor from \citet[][eq. 22]{Ahn2009}, which accounts for the interaction of LW photons redshifted into a Lyman resonance line with the gas they encounter as they travel away from their sources.
We choose to ignore the effect of the global LW background, which has been show to be sub-dominant \citep{Agarwal2012}.

While computing the LW flux, we also estimate whether a given halo is likely to be polluted by the metals produced by neighbouring haloes. For each galaxy, we compute its enrichment radius following e.g. \citet{Dijkstra2014} as the maximum distance that the gas ejected by supernovae can reach during $\Delta t \simeq 25\,\mbox{Myr}$, which is the delay between the first and last supernova for our assumed IMF. We discard such metal-enriched haloes from the list of sites eligible for SMBH seeding, but we found this to have little effect on the overall SMBH population.

Once we have computed the LW flux for each halo eligible for SMBH seeding, we compare $J_{\rm LW}$ to $J_{\rm crit}$. We seed haloes with $J_{\rm LW} < J_{\rm crit}$, with a Pop III seed with a mass $M_{\rm Pop III} = 150h^{-1}\,\Msun$, provided they have enough gas to form $M_{\star,\rm III} = 500h^{-1}\Msun$ of Pop III stars, enough to yield at least 1 BH.
For haloes above the critical LW flux, we seed them with a DCBH seed of mass $M_{\rm DCBH} = 10^{4}h^{-1}\,\Msun$ if they contain enough gas to sustain DCBH formation (which we assume to be $M_{\rm g} = 10 M_{\rm DCBH}$, but we have checked that $M_{\rm g} = M_{\rm DCBH}$ makes no difference). As we will discuss in Sect.~\Ref{sec:agn}, we do not find any DCBH seed sites in our volume, so for most for the analysis, we will discard the LW computation and assume that all SMBH seeds are Pop III seeds.

This analysis does not take into account the fact that some of the self-enriched starting haloes will already be hosting SMBHs: we correct for this by assuming that a fraction $f_{\rm seed}$ of the self-enriched haloes have in fact formed a Pop III BH. Our fiducial model assumes $f_{\rm seed} = 0.2$, but we explore values from $f_{\rm seed} = 0.1$ to $f_{\rm seed} = 1$.

\subsubsection{SMBH growth}
\label{sec:methods:agn:growth}

Once SMBH are formed, we assume they grow through two channels: gas accretion, and BH-BH mergers. For the accretion, we follow the implementation by \citet{Dayal2019} and \citet{Piana2021} in \delphi, and assume that all SMBH in haloes above a critical halo mass $M_{\rm crit}$ accrete at the Eddington limit
\begin{equation}
  \label{eq:eddington_limit}
  \dot{M}_{\rm Edd} = \frac{4\pi G \Mbh m_{\rm p}}{\epsilon_r \sigma_{\rm T} c},
\end{equation}
where \Mbh is the BH mass, $\epsilon_r = 0.1$ the radiative efficiency of the accretion flow, $G$ the gravitational constant, $m_{\rm p}$ the mass of a proton, $\sigma_{\rm T}$ the Thomson cross-section, and $c$ the speed of light.
This is in line with the results from detailed hydrodynamical simulations \citep{Dubois2015,Habouzit2017,Bower2017,Trebitsch2018, Habouzit2021} that find that that supernova feedback stunts BH growth in low-mass haloes. In particular, our critical halo mass is taken from \citet{Bower2017}:
\begin{equation}
  \label{eq:critical_halo_mass}
  M_{\rm crit} = 10^{11.25} \left(\Omega_m (1+z)^3 + \Omega_\lambda\right)^{0.125} h^{-1}\,\Msun
\end{equation}
Motivated by this, we further assume that SMBH in haloes below $M_{\rm crit}$ accrete at a small fraction $\fedd^{\rm low} = 7.5\times 10^{-5}$ of the Eddington limit.
We further assume that only a fraction of the total gas reservoir is available for accretion onto the BH, but to avoid assuming a specific gas profile in the galaxy, we leave this as a free parameter that we choose to be $f_{\bullet}^{\rm acc} = 5.5\times 10^{-4}$. This mostly makes a difference at the most massive end of the BH mass function. Overall, the accreted mass over a timestep $\Delta t$ is given by
\begin{equation}
  \label{eq:bh_accretion}
  \Mbh^{\rm acc} = (1-\epsilon_r) \min\left[\fedd \dot{M}_{\rm Edd} \Delta t, f_{\bullet}^{\rm acc} M_{\rm g}^\star\right],
\end{equation}
with $\fedd = \fedd^{\rm high}$ if $M_{\rm h} \geq M_{\rm crit}$ and $\fedd = \fedd^{\rm low}$ otherwise.

For the mergers, we assume that SMBHs merge as soon as their host halo merge. We note that this is a simplifying assumption: using cosmological simulations and modelling in details the SMBH dynamics in galaxy mergers, \citet{Volonteri2020} find that there can be a very long delay between the galaxy mergers and the actual coalescence of the two black holes. However, since we are not directly interested in measuring merger rates in this work, we follow the results of \citet{Piana2021} who found that modelling delayed BH mergers had little impact on the actual growth of the SMBH. To be conservative, we also explored a model inspired by \citet{Sassano2021} where BHs only merge rapidly during major mergers, assuming that the secondary BH in a minor merger is lost ``wandering'' in the remaining galaxy (motivated by the simulations of e.g. \citealp{Bellovary2019} and the observations of \citealp{Reines2020}), and found that it makes very little difference on the overall AGN population.

\subsubsection{AGN feedback}
\label{sec:methods:agn:feedback}

Gas accretion onto SMBH leads to an associated AGN feedback. Our implementation of AGN feedback is very similar to the way we implement SN feedback, and we follow \citet{Dayal2019}: we couple the energy released by accretion to the gas with an efficiency $f_{\bullet}^{w} = 0.3\%$ to eject gas from the halo
\begin{equation}
  \label{eq:AGN_feedback_energy}
  E_{\bullet} = f_{\bullet}^{w} \Mbh^{\rm acc} c^2
\end{equation}
It is of course possible that only a fraction of that energy is necessary to eject all the remaining gas from the halo, in which case we cap the injected energy to the energy required to lift all the remaining gas after accretion
\begin{equation}
  \label{eq:AGN_feedback_ejection}
  E_{\bullet}^{\rm ej} = \frac{1}{2} \left(M_{\rm g}^\star-\Mbh^{\rm acc}\right) v_e^2,
\end{equation}
where $v_e = \sqrt{2} v_c$ is the ejection velocity, so that the effective feedback energy is $E_{\bullet}^{\rm eff} = \min\left(E_{\bullet}, E_{\bullet}^{\rm ej}\right)$.
After this energy injection, the gas left in the halo\footnote{\label{fn:gas_post_bh}Note that here, we operate the BH growth and AGN feedback after star formation and SN feedback. We have checked that as the two processes are effectively decoupled in our model, this makes very little difference. Essentially, this is because AGN feedback is only efficient in massive haloes, that resist to their SN feedback.} is given by
\begin{equation}
  \label{eq:AGN_gas_remaining}
  M_{\rm g}^{\bullet} = \left(M_{\rm g}^\star - M_{\bullet}^{\rm acc}\right) \left(1 - \frac{E_{\bullet}^{\rm eff}}{E_{\bullet}^{\rm ej}}\right).
\end{equation}

\subsubsection{AGN ionizing emissivity}
\label{sec:methods:agn:radiation}

Associated to the release of energy through AGN feedback, SMBH affect their environment by releasing radiation. We assign an ionizing luminosity to each AGN in our simulation following the formalism of \citet{Volonteri2017} as already implemented in \citet{Dayal2020}. Each AGN is assigned an SED that depends on the mass of the SMBH and its Eddington ratio, following the model of \citet{Done2012}. The peak of the SED is computed using the method of \citet{Thomas2016}, while the global shape of the spectrum is assumed to follow the functional form used in \textsc{Cloudy} \citep{Ferland2013}. We integrate the SED above $13.6\,\mbox{eV}$ to compute the ionizing luminosity and the mean energy of the ionizing radiation, including a correction for secondary resulting from hard photons assuming the maximal possible contribution i.e. that they propagate in fully neutral hydrogen and that $39\%$ of their energy is available for secondary ionizations, \citep{Shull1985, Madau2017}. The resulting luminosity is shown in Appendix A1 of \citet{Dayal2020}.
From this, we derive the ionizing photons production rate from each AGN, that we use as input for the reionization module of \astraeus.

To estimate the contribution of AGN ionizing radiation to the UV background, we further need to estimate which fraction $\fescAGN$ of the photons escape the galaxy. In this work, we explore four different models.
\begin{enumerate}
\item First, we use a model in which we assume that $\fescAGN = 1$. This is in line e.g. with the findings of \citet{Cristiani2016}, who found on average a high AGN escape fraction in their quasar sample at $3.6 < z < 4.0$. We note that this is a fairly extreme model, which focuses on quasars rather than more normal AGN. For instance the simulations of \citet{Trebitsch2018,Trebitsch2021} or the observations of \citet{Micheva2017} suggests a lower \fescAGN for less luminous AGN. Nevertheless, this will let us estimate the maximum contribution from AGN to reionization allowed by our model.
\item Second, we assume a less extreme scenario in which the escape fraction is essentially set by (one minus) the obscured fraction of AGN. We use the redshift-independent obscured fraction derived by \citet{Merloni2014}:
\begin{equation}
  \label{eq:fobs_merloni}
  f_{\rm obs} = 0.56 + \frac{1}{\pi}\arctan\left(\frac{43.89 - \log_{10}(L_{\rm X})}{0.46}\right)  
\end{equation}
where $L_{\rm X}$ is the X-ray luminosity of the AGN in $\mbox{erg}\,\mbox{s}^{-1}$ that we estimate from the bolometric luminosity $L_{\rm bol}$ as  $L_{\rm X} = L_{\rm bol}/K_{\rm X}$ using the bolometric correction from \citet{Duras2020}:
\begin{equation}
  \label{eq:kbol_duras}
  K_{\rm X} = 10.96 \left(1 + \left(\frac{\log_{10}(L_{\rm bol}/L_{\sun})}{11.93}\right)^{17.79}\right),
\end{equation}
with $L_{\sun}$ the Solar luminosity.
For each AGN, we use the $f_{\rm obs}$ corresponding to its luminosity to randomly draw whether it is obscured or not. We then set the AGN escape fraction to $\fescAGN = 1$ for unobscured AGN and $\fescAGN = 0$ otherwise. Our choice of using the \citet{Merloni2014} obscuration fraction, derived at $z \lesssim 3.5$, rather than e.g. the \citet{Vito2018} estimate of the obscuration fraction at $3 < z < 6$ is motivated by the chosen definition of `obscured AGN': since we are motivated by the escape of UV and ionizing radiation, the optical classification of \citet{Merloni2014} is more relevant than an X-ray classification that will be more indicative of whether an AGN is heavily obscured or not.
\item Our third model is a variation on the previous one, where we assume that all AGN have an escape fraction equal to the unobscured fraction: $\fescAGN = 1 - f_{\rm obs}$. Assuming no radiative feedback, this essentially yields the same ionizing budget as the previous model, but with a different spatial distribution.
\item Finally, we explore a model where we assume that \fescAGN is identical to the \fescstar of the galaxy population.
\end{enumerate}

\section{Calibration and AGN population}
\label{sec:agn}

We now proceed to calibrate our AGN model against observations to fix the parameters of our fiducial model, and discuss the resulting AGN population in relation with the host galaxy population.

\begin{figure}
  \centering
  \includegraphics[width=\columnwidth]{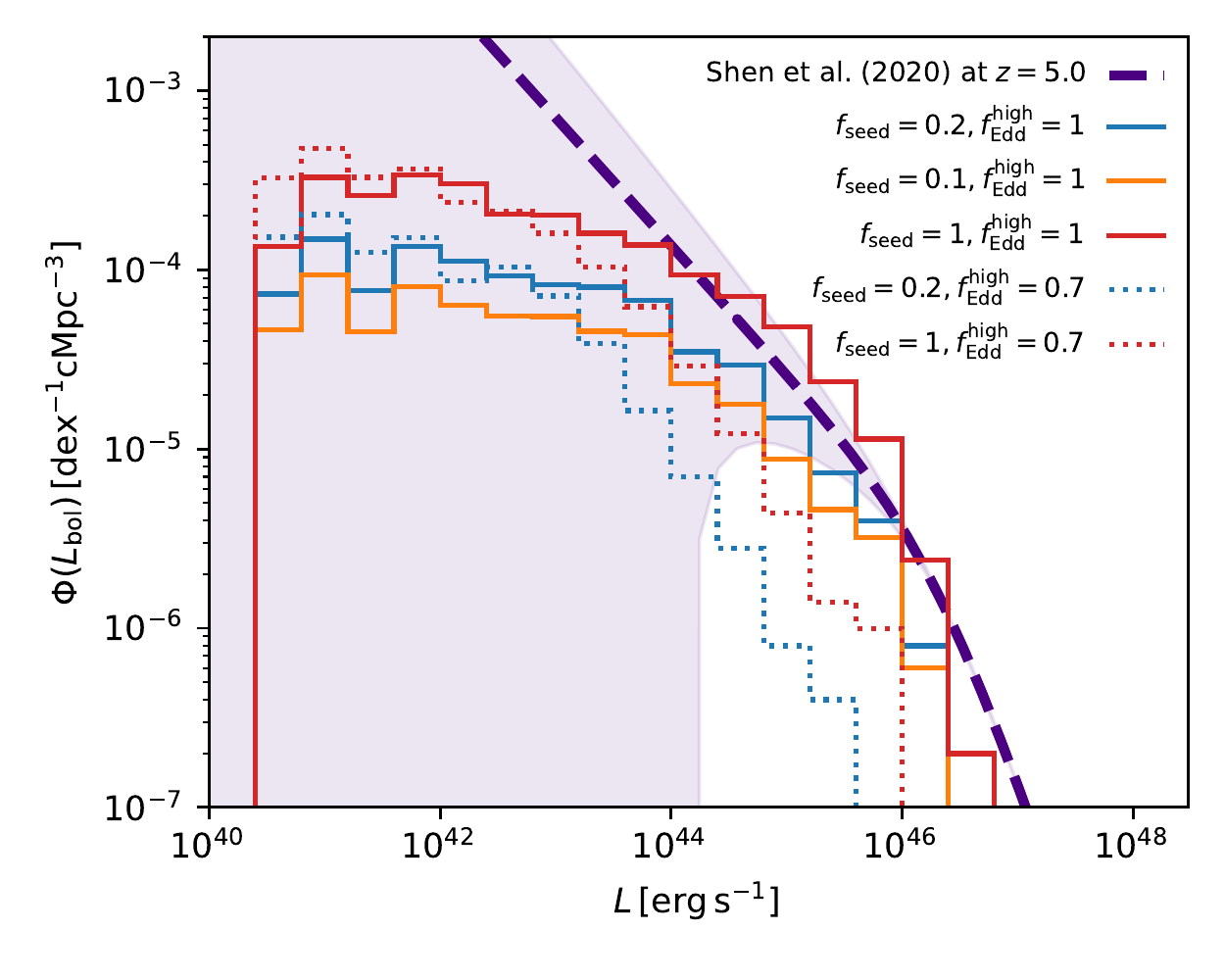}
  \includegraphics[width=\columnwidth]{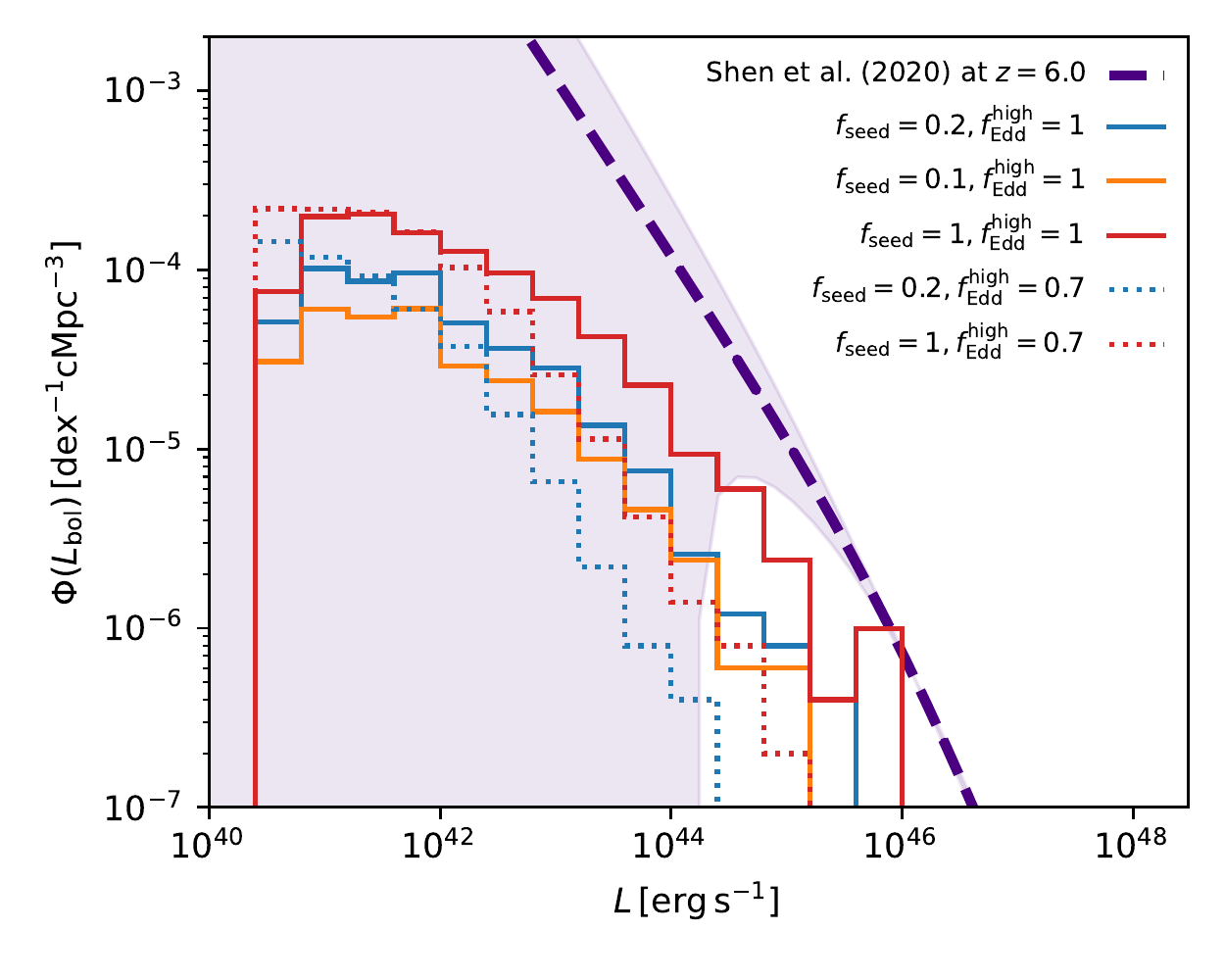}
    \includegraphics[width=\columnwidth]{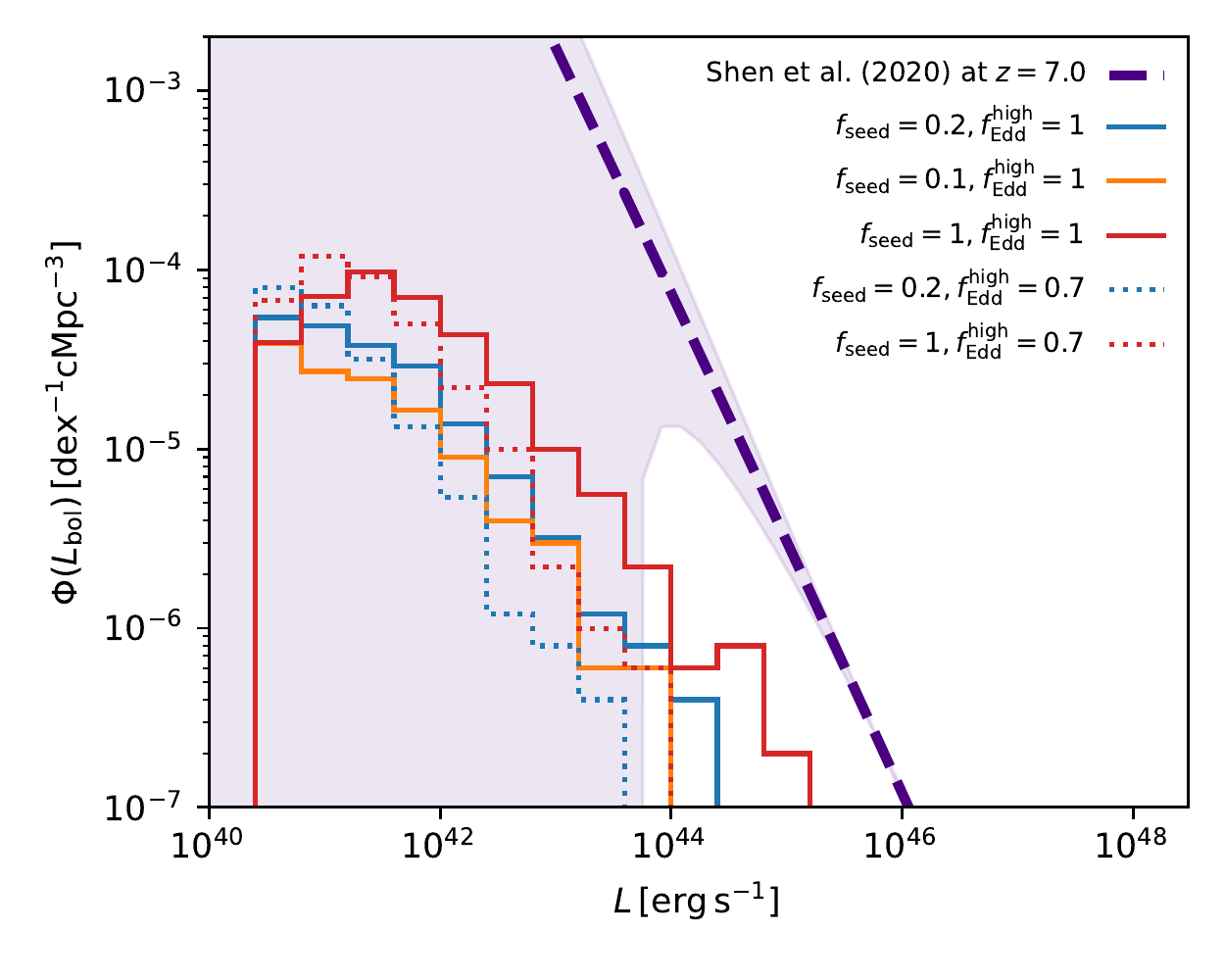}\vspace{-1em}
  \caption{AGN bolometric luminosity functions at $z=5$ (upper panel), $z=6$ (middle panel), and $z=7$ (lower panel) for runs with $f_{\rm seed} = 0.1$ (orange), $0.2$ (blue), and $1$ (red) for two different Eddington ratios in high-mass haloes: $\fedd^{\rm high} = 1$ (solid lines) and $\fedd^{\rm high} = 0.7$ (dotted lines). The thick purple dashed line is the global evolution fit from \citet{Shen2020}, with the shaded area corresponding to the (propagated) uncertainty on their best-fit parameters. The parameters that reproduce the best the observed LF are $f_{\rm seed} = 0.2$ and $\fedd^{\rm high} = 1$.}
  \label{fig:AGN_LbolLF}
\end{figure}

\subsection{Model calibration}
\label{sec:agn:calbiration}

We have chosen the AGN bolometric LF as our main constraint for our AGN model. While it is not straightforward to infer from observations, this is the most direct outcome of the model that includes the effects of both the seeding and the growth prescription, and can be inferred from our model without having to invoke any obscuration prescription. 
From our simulations, we compute the bolometric luminosity of each AGN as
\begin{equation}
  \label{eq:AGN_Lbol}
  L_{\rm bol} = \epsilon_r \dot{\Mbh} c^2 = \epsilon_r \frac{\Mbh^{\rm acc}}{\Delta t} c^2
\end{equation}

Our two main free parameters for this calibration are the fraction of pre-enriched haloes hosting Pop III BH, $f_{\rm seed}$, and the Eddington ratio in high-mass haloes, $\fedd^{\rm high}$. We have checked that changing $f_{\bullet}^{\rm acc}$ has little impact on the overall shape of the bolometric LF, and only affects the growth of the most massive BHs at $z \lesssim 5$. Similarly, we have verified that changing $\fedd^{\rm low}$ to a value 100 times higher makes virtually no difference on the AGN LF in the regime where observational constraints exist.

The upper panel of Fig.~\ref{fig:AGN_LbolLF} shows the bolometric LF at $z=5$ for runs with $f_{\rm seed} = 0.1, 0.2, 1$ and $\fedd^{\rm high} = 1$ and $0.7$.
As expected, the models with a lower \fedd tend to yield a lower number density of bright AGN (or rather, at fixed number density, the AGN are fainter with lower \fedd), while the overall normalisation is set by $f_{\rm seed}$.
In our model, only the growth of the most massive BHs in the most massive haloes is limited by the gas supply ($f_{\bullet}^{\rm acc} M_{\rm g}^\star$ in eq.~\ref{eq:bh_accretion}). For the runs with lower \fedd, BHs will enter this regime at higher mass and will therefore accrete most of the time at $\fedd$. In that context, a lower \fedd naturally yields a slower growth of the BH, and therefore a lower luminosity at fixed number density.
We use as our main constraint the ``global fit'' model from recent quasar bolometric LF from \citet{Shen2020}, which is based on a compilation of observations at $z = 0-7$, and includes the contributions of both obscured and unobscured quasars. At $z = 5$, the observational data points constraining the \citet{Shen2020} fit range above luminosities of $L_{\rm bol} \gtrsim 10^{44}\,\mbox{erg}\,\mbox{s}^{-1}$.

We find that the models that best reproduce the observed LF at $z=5$ in the range probed by observations are those with $f_{\rm seed} = 0.1 - 0.2$ and $\fedd^{\rm high} = 1$. The model with $f_{\rm seed} = 0.1$ tends to give a slightly better match to the brighter end of the LF, but performs slightly worse at the fainter end. By comparison, when lowering $\fedd^{\rm high}$, even the model where 100\% of the pre-enriched haloes are hosting a Pop III BH seed tends to produce too few AGN compared to the observations.

We show the same bolometric LF at $z=6$ ($z=7$) in the central (lower) panel of Fig.~\ref{fig:AGN_LbolLF}. We find that all our models tend to under-predict the number of AGN at $z > 6$, with only the most extreme one ($f_{\rm seed} = 1$ and $\fedd^{\rm high} = 1$) reaching the observed LF at $z=6$ at its bright end. The observations at $z \sim 6$ only constrain AGN brighter than $L_{\rm bol} \gtrsim 10^{45}\,\mbox{erg}\,\mbox{s}^{-1}$, so this model would in principle be an acceptable match in that luminosity regime. However, given that it overestimates the LF at lower $z$, we only regard this model as a ``maximal'' case for the rest of this work.
At even higher $z$, our models fall short of the \citet{Shen2020} bolometric LF, but we stress that this comes from the extrapolation of their fit rather than from data. Overall, we find that the bolometric LF grows faster in our model than in the (extrapolation) of observations, with a very low number density of bright AGN at high redshift.

\begin{figure}
  \centering
  \includegraphics[width=\columnwidth]{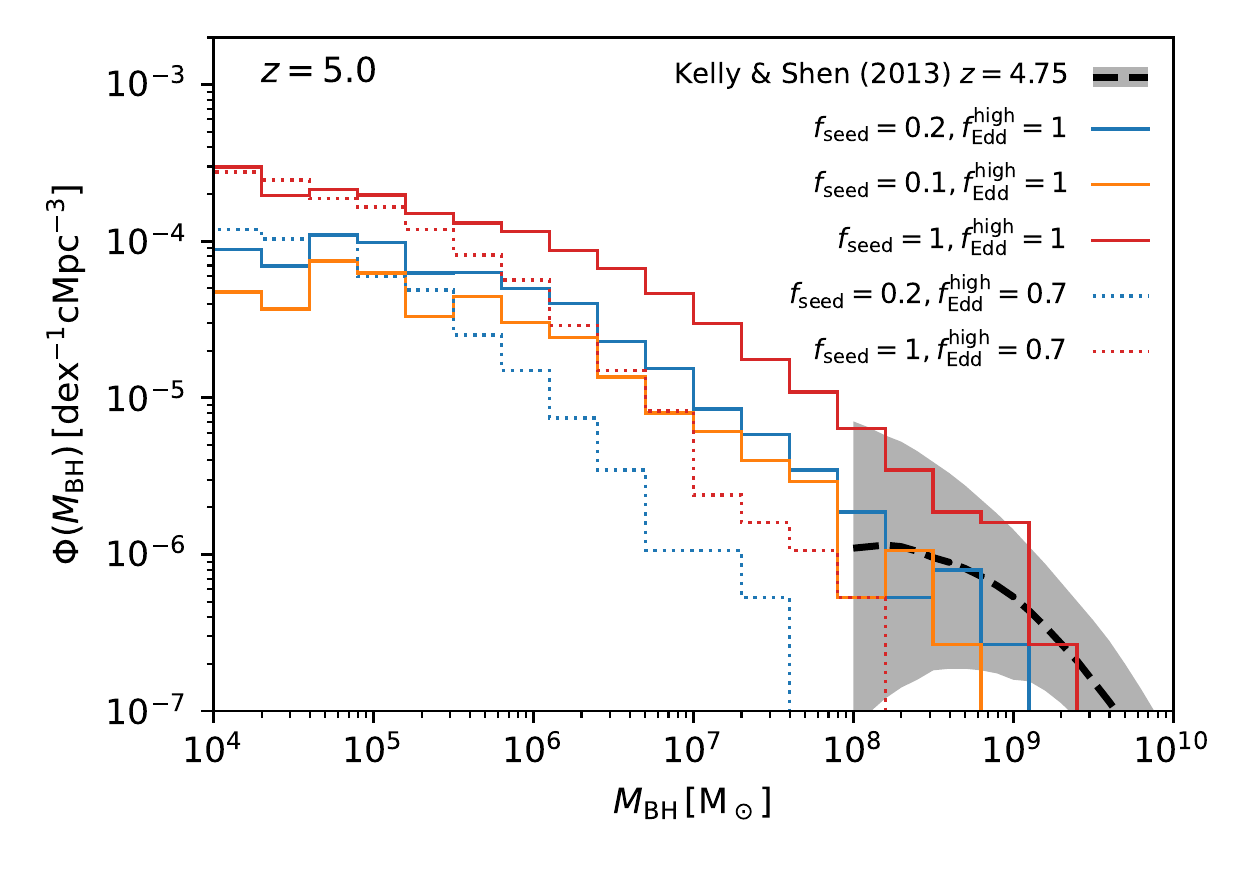}
  \includegraphics[width=\columnwidth]{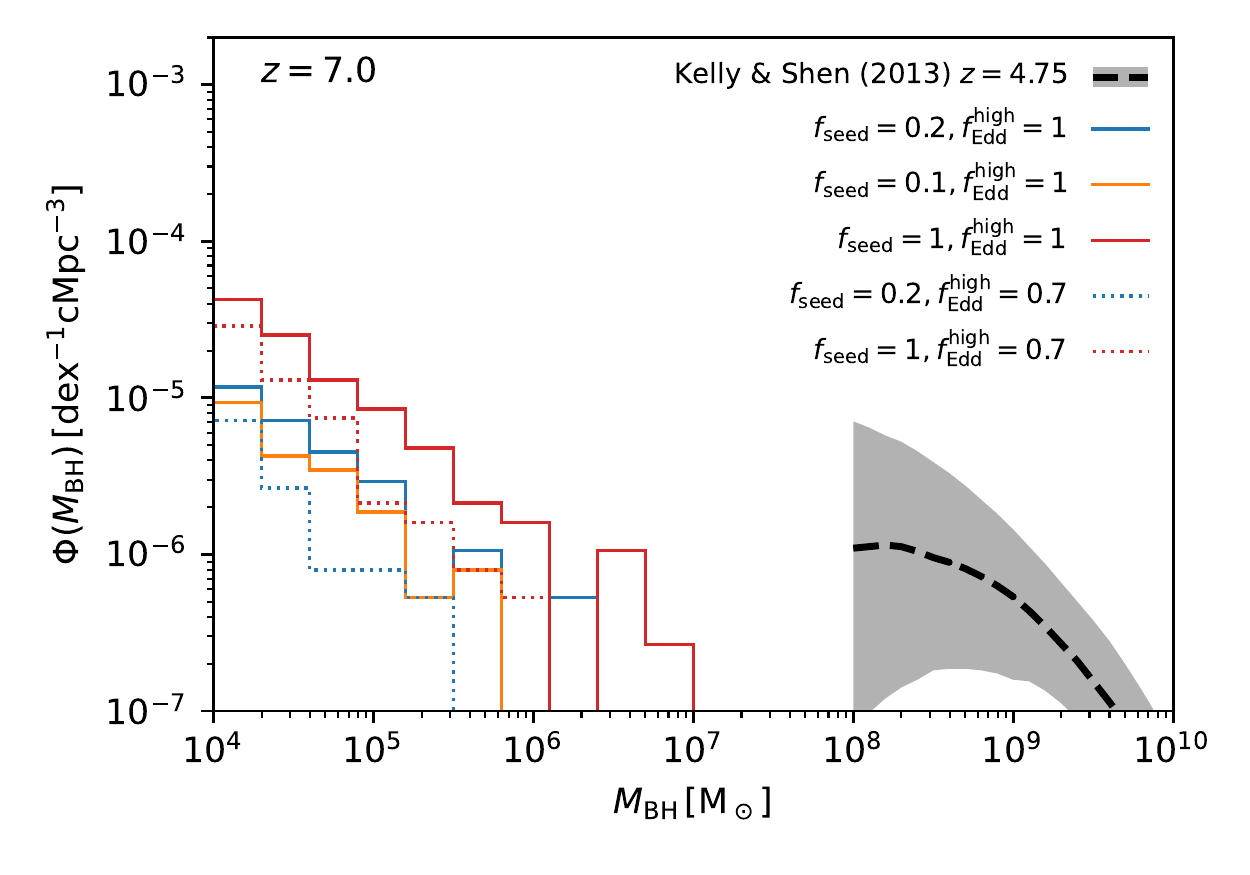}
  \caption{BH mass function at $z=5.0$ (top) and $z=7.0$ (bottom) for the five models shown in Fig.~\ref{fig:AGN_LbolLF}, compared to the \citet{Kelly2013} observations at $z=4.75$ (dashed black line for the median and grey area for the 16\textsuperscript{th}-84\textsuperscript{th} percentile range). The shape of the BH mass function is mostly affected by the choice of $\fedd^{\rm high}$ at the massive end, while the normalisation follows $f_{\rm seed}$.}
  \label{fig:BHMF}
\end{figure}

\begin{figure*}
  \centering
  \includegraphics[width=\linewidth]{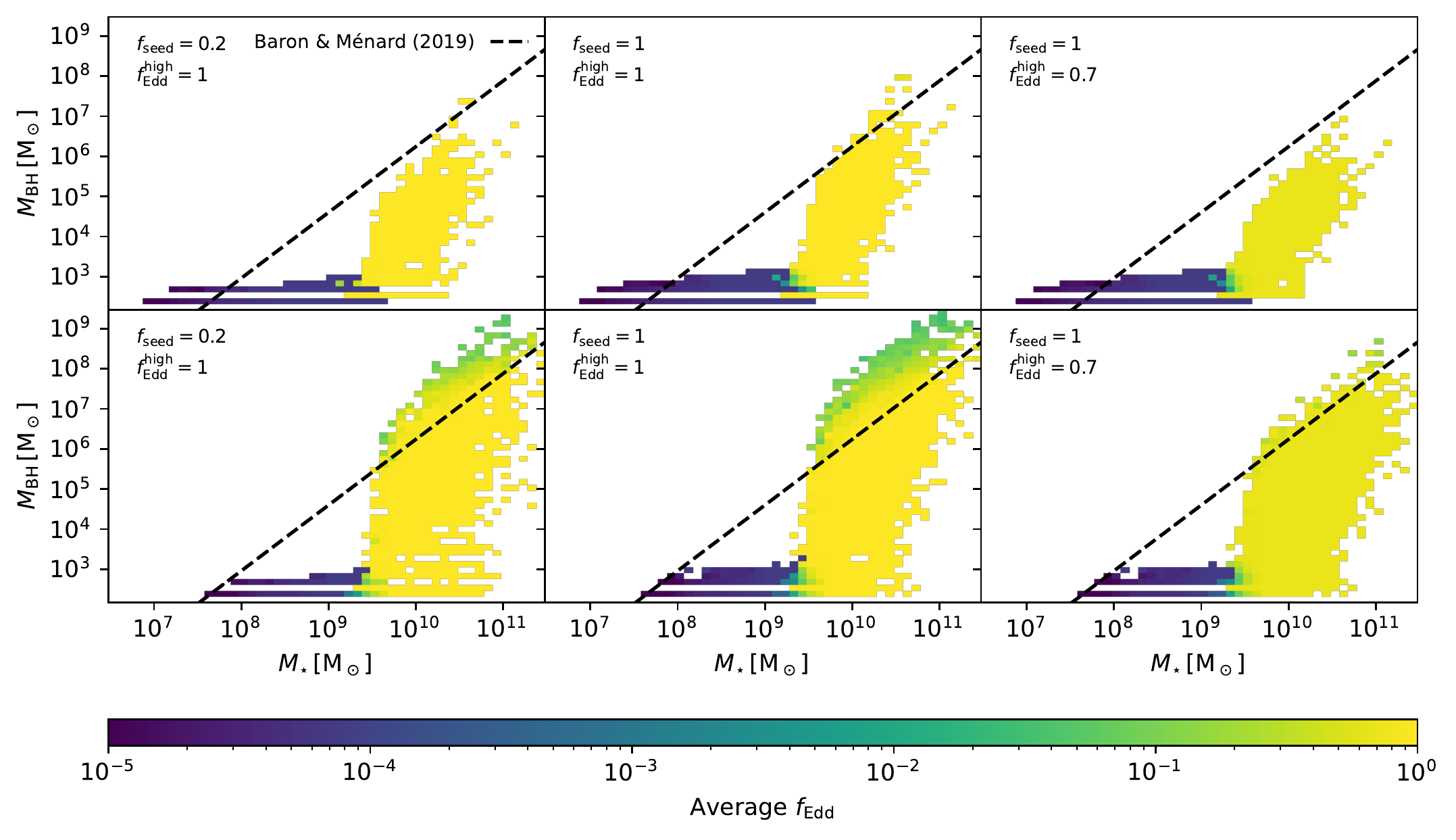}
  \caption{$\Mbh-\Mstar$ relation for models with $\fedd^{\rm high} = 1$ and $f_{\rm seed} = 0.1$ (left), $\fedd^{\rm high} = 1$ and $f_{\rm seed} = 1$ (centre), and $\fedd^{\rm high} = 0.7$ and $f_{\rm seed} = 1$ (right), at $z=6$ (top) and $z=4.5$ (bottom). The colour indicates the average \fedd in each bin. The \citet{Baron2019} $z\simeq 0$ relation is shown as a dashed line.}
  \label{fig:MstarMBH}
\end{figure*}

\subsection{Black hole masses}
\label{sec:agn:masses}

We show in Fig.~\ref{fig:BHMF} the BH mass function at $z = 5$ (top) and $z=7$ (bottom) for the five models used to calibrate our parameters, using the same colours and line styles as in Fig.~\ref{fig:AGN_LbolLF}. For all the models with $\fedd^{\rm high} = 1$, the global shape of the mass function is unchanged, with the normalisation following $f_{\rm seed}$. This is to be expected (see also Sect.~\ref{sec:agn:occupation}) since our choice of $f_{\rm seed}$ affects haloes of all masses equally.
By comparison, reducing $\fedd^{\rm high}$ to $0.7$ predominantly affects the massive end of the mass function. Since BHs only accrete at $\fedd^{\rm high}$ if they are in haloes massive enough, this suggests that only BHs with masses $\Mbh \gtrsim 10^6\,\Msun$ tend to live in massive haloes.
We also show the $z=4.85$ BH mass function derived from observations of high-$z$ quasars by \citet{Kelly2013} as a dashed black line, with the $16^{\rm th}$ and $84^{\rm th}$ percentiles indicated by the grey area. Comparing the $z=5$ mass functions, the models with a low $\fedd^{\rm high}$ are disfavoured by the comparison to observations: they all under-predict the number density of BHs with masses above $\Mbh \gtrsim 10^8\,\Msun$. By contrast, all our models with $\fedd^{\rm high} = 1$ are in reasonable agreement with the \citet{Kelly2013} mass function, and the model with $f_{\rm seed} = 0.2$ provides the best match to the observations.
By comparing the $z=5$ and $z=7$ mass functions, we can see that our model doesn't predict any quasar-like extremely massive BH at $z=7$, consistent with the volume we are probing. We find that our most massive BHs grow late, between $z=7$ and $z=5$.

We explore this further in Fig.~\ref{fig:MstarMBH}, where we show the relation between BH and galaxy stellar mass at $z=6$ (top row) and $z=4.5$ (bottom row) for the models with $\fedd^{\rm high} = 1$ and $f_{\rm seed} = 0.1$ (left panels), $\fedd^{\rm high} = 1$ and $f_{\rm seed} = 1$ (central panels), and $\fedd^{\rm high} = 0.7$ and $f_{\rm seed} = 1$ (right panels). The colour coding indicate the average Eddington ratio \fedd in each mass bin. The dashed line on all panels shows the relation derived by \citet{Baron2019} at $z \simeq 0$, extrapolated down to arbitrarily low stellar masses.
As expected from our modelling choice of $\fedd^{\rm low} \ll 1$, the growth of the BHs in all our models is completely stunted in galaxies with masses below $\Mstar\lesssim 10^{9.5}\,\Msun$ (as indicated by the very low average \fedd). Once galaxies reach that mass, BHs grow efficiently at the Eddington rate and reach masses in good agreement with the $z \sim 0$ expectations from their host stellar mass. The slope of the BH-to-stellar mass relation in our model becomes shallower at the very high-mass ($\Mstar \gtrsim 10^{10.5}\,\Msun$) in all of our models, indicating that BH growth is no longer proceeding at $\fedd^{\rm high}$, but instead hindered by the amount of gas available: the second term in Eq.~\ref{eq:bh_accretion} becomes the limiting factor in estimating the accretion rate.
This is more obvious on the lower panel, at $z = 4.5$: at fixed galaxy mass, the most massive BHs grow at less than $10\%$ of the Eddington rate.
In the model with $f_{\rm seed} = 0.2$, very few BHs are already on the local $\Mstar - \Mbh$ relation at $z = 6$. As we will see in Sect.~\ref{sec:agn:occupation}), this is not mainly caused by a difference in occupation fraction at high mass. Instead, this is caused by the fact that with fewer seed, the contribution of mergers to BH growth is more limited in this model.

Comparing the central and right panels of Fig.~\ref{fig:MstarMBH}, we can see that the main effect of limiting the Eddington ratio in high-mass haloes is to make the overall shape of the $\Mstar - \Mbh$ relation shallower at $\Mstar\gtrsim 10^{9.5}\,\Msun$. This directly comes from the fact that a lower $\fedd^{\rm high}$ leads to a slower growth of the BHs. Since in our model AGN feedback has little effect on star formation (apart at the highest masses), the stellar mass is virtually unchanged when varying $\fedd^{\rm high}$, therefore resulting in a shallower slope. While it seems that at $z = 4.5$, the $\fedd^{\rm high} = 0.7$ model provides a better fit to the local $\Mstar - \Mbh$ relation, we refrain from putting too much weight on this: our growth model assumes a constant $\fedd^{\rm high}$, while observations at lower redshift require that the average $\fedd$ decreases with time \citep[e.g.][]{Kelly2013}. Similarly, because the $z \gtrsim 5$ constraints on the Eddington ratio distribution are extremely sparse, we have chosen to assume a single value instead of assuming a wider distribution as in e.g. \citet{Shankar2013, Volonteri2017}. Because of this, we are missing the population of BHs with milder growth, and our models with $\fedd^{\rm high}$ here are to be taken as maximal cases for the growth of Pop III seeds, especially since we assume that BH merge instantaneously when their host haloes merge.

\subsection{Occupation fraction}
\label{sec:agn:occupation}

\begin{figure}
  \centering
  \includegraphics[width=\columnwidth]{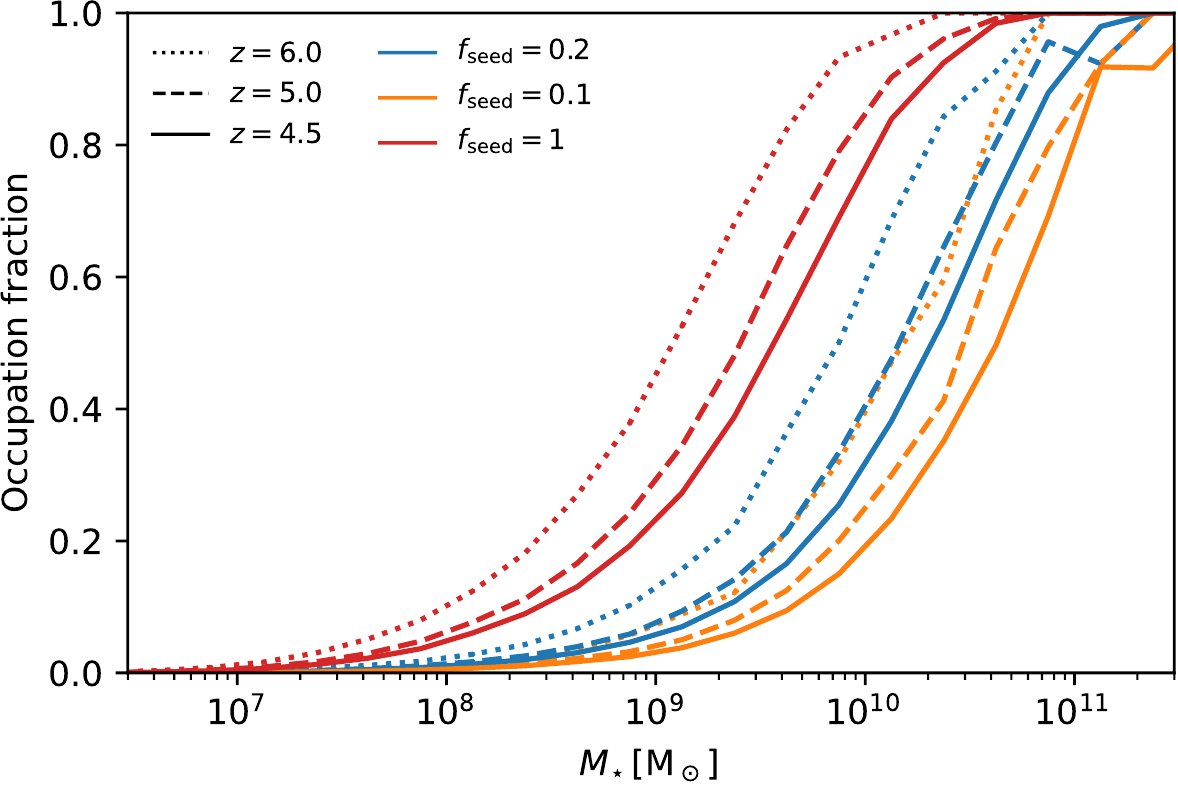}
  \caption{BH occupation fraction at $z=6.0$ (dotted line), $z=5.0$ (dashed line) and $z=4.5$ (solid line) for the three different $f_{\rm seed}$ using the same colours as in Fig.~\ref{fig:AGN_LbolLF}. At fixed stellar mass, the occupation fraction decreases with $z$ because black holes stop forming at $z \gtrsim 12$, while galaxies keep growing.}
  \label{fig:OccFrac}
\end{figure}

Observationally, not all galaxies contain active BHs. While this comes in part from the fact that not all BHs are actively accreting matter, theoretical models of BH formation do not predict that BH seeds are ubiquitous \citep[see e.g.][]{Volonteri2010, Inayoshi2020}, and in particular most scenarios require extremely metal-poor gas for BH seeds to form. Different seeding models predict different occupation fractions for the seeds, and therefore for the black holes growing from these seeds.
In our model, as the IGM metallicity increases over time, the metallicity of newly identified haloes also increases, so that at $z \lesssim 12$ no new BHs are formed.
We show the occupation fractions of all BHs as a function of stellar mass of the host galaxy in Fig.~\ref{fig:OccFrac} for the three models with $\fedd^{\rm high} = 1$ and $f_{\rm seed} = 0.1$ (orange), $0.2$ (blue) and $1$ (red). For each model, the different lines indicate different redshifts: $z=6.0$ (dotted line), $z=5.0$ (dashed line) and $z=4.5$ (solid line).
As expected, the occupation fraction is higher in the models with a higher $f_{\rm seed}$: more haloes are initially hosting a BH seed, so that more haloes will be hosting BHs at later time.
At fixed occupation fraction, the corresponding host mass increases with decreasing redshift. This can be understood easily as no new BHs are formed at $z \lesssim 12$. Galaxies will steadily grow whether or not they are hosting a BH, so that the curves are all moving towards higher masses as $z$ decreases.

Comparing our results with the cosmological simulation of \citet{Habouzit2017}, which explicitly focuses on following the formation of Pop III seeds, it seems that we systematically under-predict the occupation fraction. Only our extreme model with $f_{\rm seed} = 1$ appears to be marginally in agreement with their simulation, and only when comparing with their ``inefficient SN feedback'' models. Their much higher occupation fraction result from the fact that our BH seeds stop forming at much higher $z$ than in \citet{Habouzit2017}, where the ISM metallicity is tracked down to a resolution of $\simeq 75\,\mbox{pc}$. This means that new haloes will form seeds down to much lower $z$ in their simulation, and so the overall occupation fraction will be higher.
Interestingly, they find that this model over-predicts the AGN bolometric LF significantly more than we do even for our $f_{\rm seed} = 1$ model. This high apparent LF can be reconciled with observations by assuming a duty-cycle of order $10-20\%$, while our model implicitly assumes a duty-cycle of $100\%$. 
This highlights the sensitivity of BH and AGN models to the ISM prescription: detailed simulations such as those of \citet{Trebitsch2019} have found that the actual BH growth duty-cycle is higher than the observed AGN duty-cycle.
Nevertheless, the good agreement between our models and both the observed AGN bolometric LF and BH mass function validate their use to study the AGN contribution to reionization.

\subsection{UV emission}
\label{sec:agn:uv}
We now turn to the UV luminosity produced by our AGN population. For this, we use the same bolometric correction as in \citet{Shen2020} to estimate the UV luminosity of our AGN (but we checked that this had little impact on our results, comparing e.g. with the correction from \citealt{Runnoe2012}).

\subsubsection{AGN UV luminosity function}
\label{sec:agn:uv:agn_uvlf}

\begin{figure}
  \centering
  \includegraphics[width=\columnwidth]{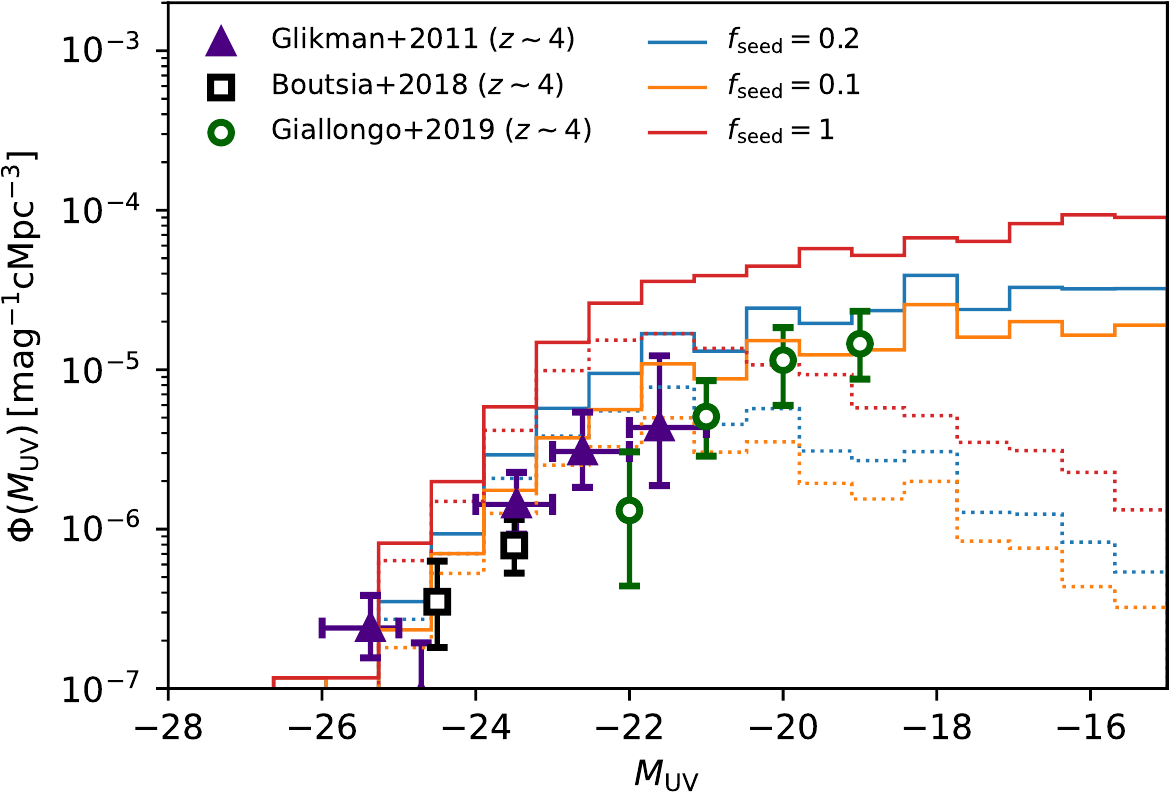}\\
  \includegraphics[width=\columnwidth]{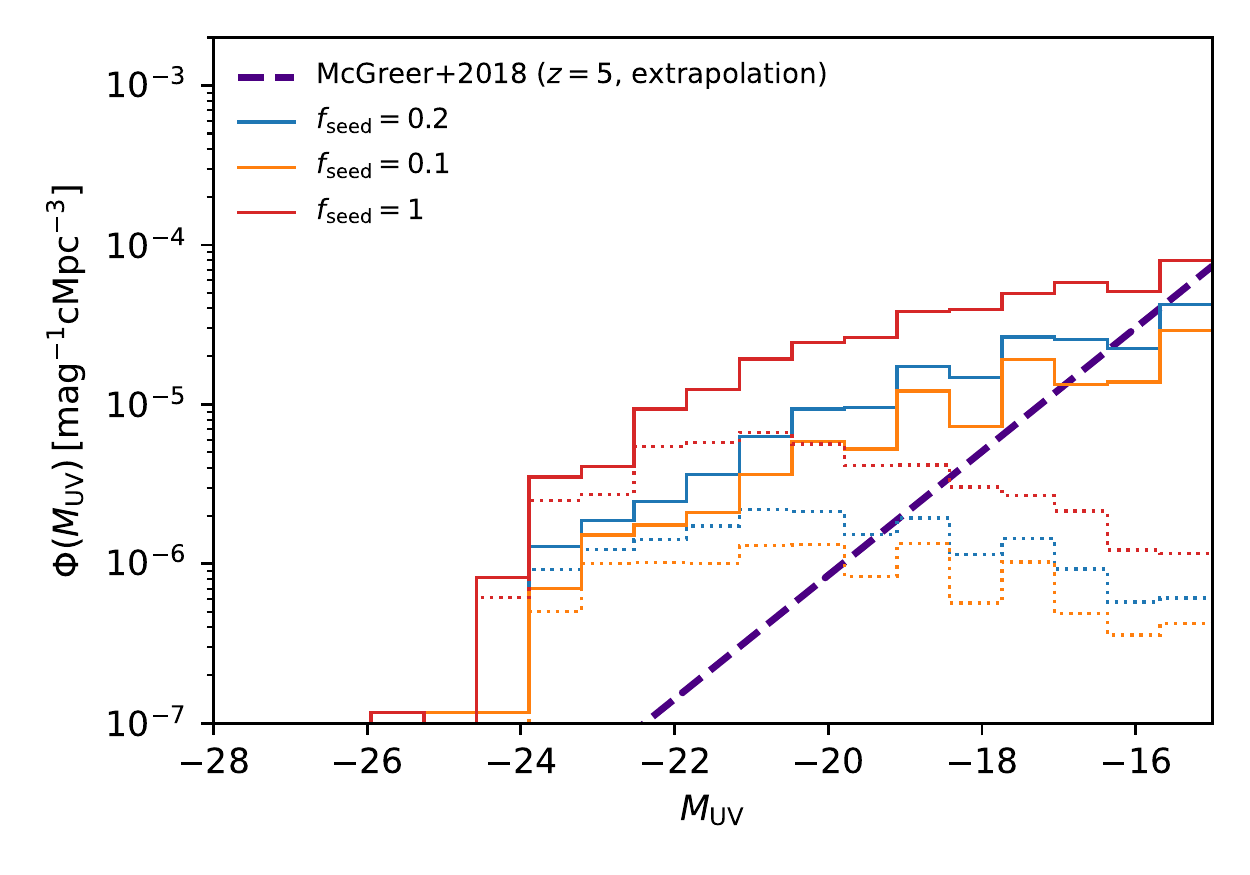}\\
  \includegraphics[width=\columnwidth]{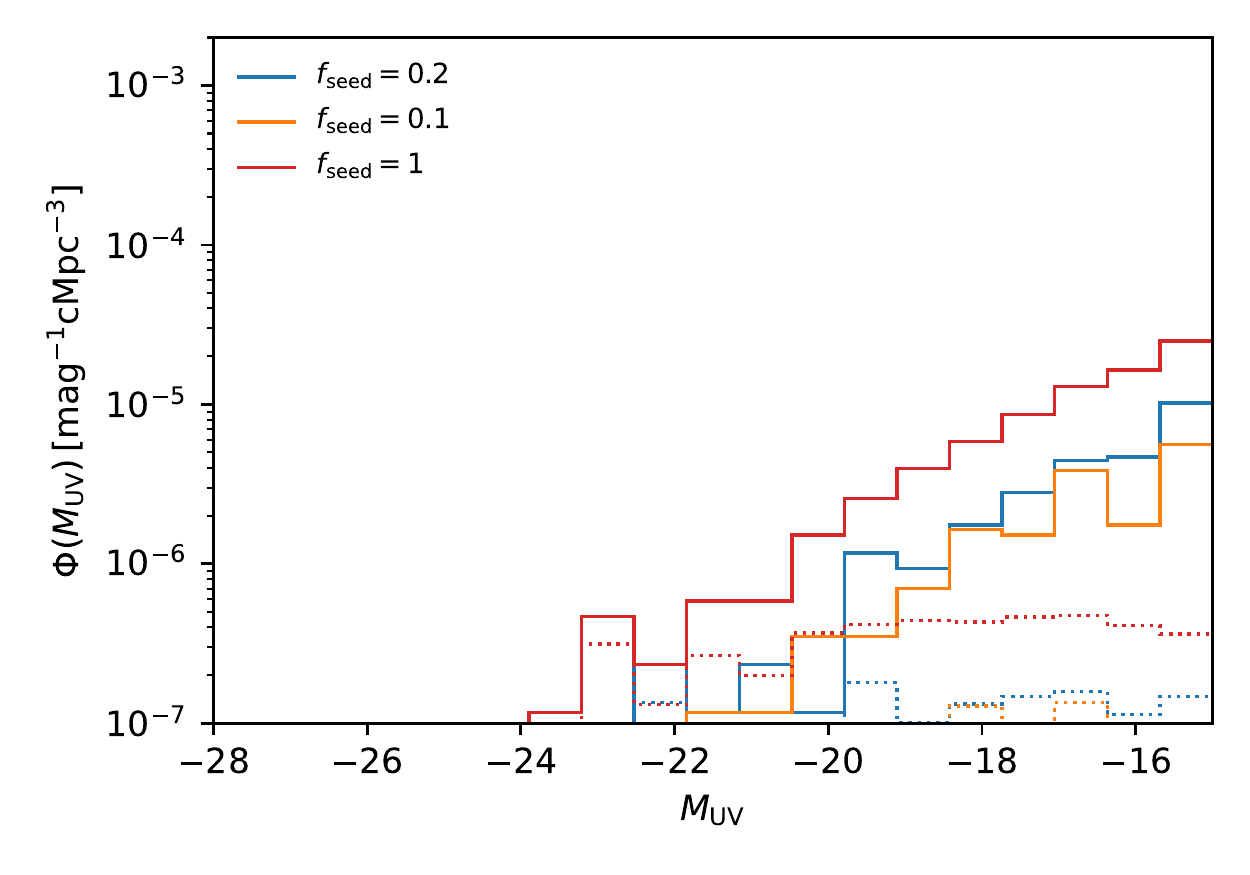}\\
  \caption{AGN UV luminosity functions at $z=4.5$ (top), $z=5$ (middle) and $z=6$ (bottom) for runs with $f_{\rm seed} = 0.1$ (orange), $0.2$ (blue), and $1$ (red), compared at $z=4$ to  observations by \citet[][purple triangles]{Glikman2011}, \citet[][black squares]{Boutsia2018} and \citet[][green circles]{Giallongo2019} (top), and at $z=5$ to an extrapolation of the luminosity function of \citet[][purple dashed line]{McGreer2018} (middle). The dotted lines correspond to the UV LF including the obscuration from \citet{Merloni2014}.}
  \label{fig:AGN_UVLF}
\end{figure}
We show in the top panel of Fig.~\ref{fig:AGN_UVLF} the intrinsic AGN UV LF at the end of the simulation ($z=4.5$) for the two best-fit models ($f_{\rm seed} = 0.1 - 0.2$) as well as our more extreme model ($f_{\rm seed} = 1$) using the same colour scheme as in Fig.~\ref{fig:AGN_LbolLF}. We include a correction for the obscured AGN following \citet{Merloni2014}, using the prescription described with Eq.~\ref{eq:fobs_merloni} and \ref{eq:kbol_duras}.
By comparison, observational estimates of the UVLF at $z \sim 4$ are shown as purple triangles \citep{Glikman2011}, black squares \citep{Boutsia2018} and green circles \citep{Giallongo2019}. 

Our two best fit models show a trend that is overall consistent with the observed AGN UV LF, with a slightly better match obtained for the $f_{\rm seed} = 0.2$ model when taking obscuration into account. 
The extreme, $f_{\rm seed} = 1$ model overshoots the observed LF at $M_{\rm UV}$ brighter than $-22$ even when including the effect of obscuration. The obscured fraction for \citet{Merloni2014} is only effectively constrained at X-ray luminosities $L_X \gtrsim 10^{43}\,\mbox{erg}\,\mbox{s}^{-1}$ corresponding to $M_{\rm UV} \simeq -18$, well below the apparent turn-over in our LF: at face value, this suggest that we under-estimate the faint-end of the AGN UV LF when folding in the effect of dust, compared to the results of \citet{Giallongo2019}. We note however that their LF was derived under the assumption that the observed UV was predominantly coming from an AGN component: this effectively corresponds to assuming that all their AGN are unobscured. It is therefore more reasonable to compare the faint end of the AGN UV LF from \citet{Giallongo2019} to our unobscured AGN UV LF. In that case, our two standard models are in good agreement with their observed LF.

In the middle and lower panel, we show the same AGN UV LFs at $z=5$ and $z=6$, respectively. As expected from the evolution of the bolometric LF discussed in Sect.~\ref{sec:agn:calbiration}, it appears clearly that the number density of UV-bright AGN drops significantly at higher redshifts. At $z=5$, we compare our results to an extrapolation of the $z\sim 5$ LF from \citet{McGreer2018} (shown as a purple dashed line): the data stops at a number density of $\Phi \lesssim 10^{-7}\,\mbox{mag}^{-1}\mbox{cMpc}^{-3}$, corresponding to just under one object in our simulation volume. For the same reason, we note that we cannot directly compare our results to observational determinations of the AGN UV LF at $z \gtrsim 6$, such as those resulting from the SHELLQs survey \citep{Matsuoka2018} because they probe number densities too low to be sampled in our cosmological volume. Nevertheless, we tentatively find a number density of AGN larger than suggested by \citet{McGreer2018}, even after accounting for obscuration. One possible explanation could be that we under-estimate the obscuration for these objects: consistent with the observations of \citet{Circosta2019}, \citet{Trebitsch2019} found that the ISM can significantly contribute to the AGN obscuration in massive high-$z$ galaxies. In any case, Fig.~\ref{fig:AGN_UVLF} points towards a rapid evolution of the AGN UV LFs at $z \gtrsim 5$.

\subsubsection{Combined UV luminosity function}
\label{sec:agn:uv:combined}

\begin{figure}
  \centering
  \includegraphics[width=\columnwidth]{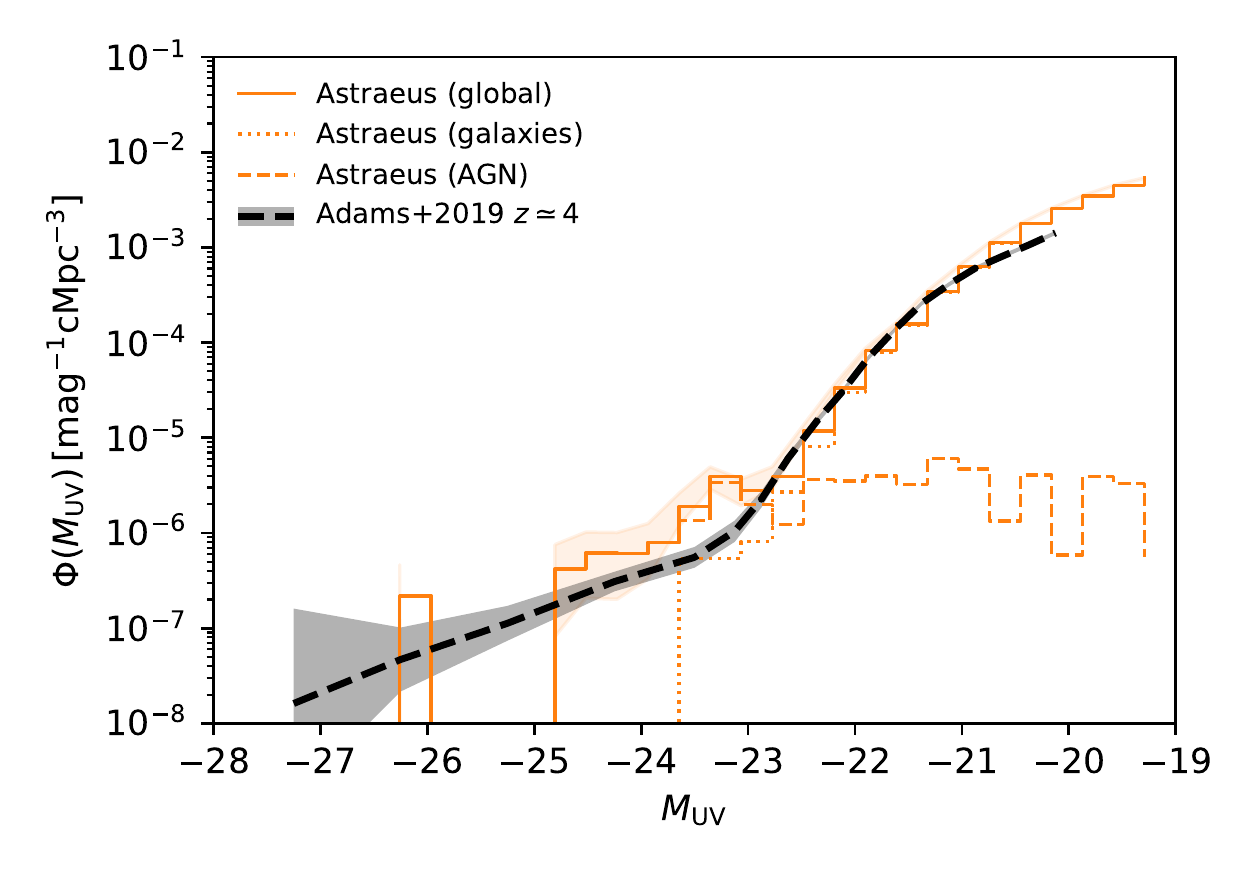}
  \caption{Combined AGN+galaxy UV LF at $z=4.5$ for the $f_{\rm seed} = 0.1, \fedd^{\rm high} = 1$ model (orange solid line), and the individual AGN (dashed line) and galaxy (dotted line) components, after including attenuation from dust. The dashed black line and grey area are the combined UV LF from \citet{Adams2020} at $z \simeq 4$. In good agreement with observations, our AGN start to dominate the UV LF around $\MUV \simeq -23$.}
  \label{fig:AGN_UVLF_global}
\end{figure}

In the past few years, multiple groups \citep[e.g.][]{Ono2018, Stevans2018, Adams2020, Harikane2022} have studied in detail the intersection between the bright-end of the galaxy UV LF and the faint-end of the AGN UV LF. Since \astraeus models star-forming galaxies and AGN together, we can estimate the UV luminosity function of all sources in our simulation. We show in Fig.~\ref{fig:AGN_UVLF_global} the results from the $f_{\rm seed} = 0.1, \fedd^{\rm high} = 1$ model in orange, with the solid line corresponding to the combined LF, the dashed (dotted) line being the AGN (galaxy) contribution. The galaxy contribution includes dust attenuation, while the AGN contribution takes obscuration into account, as in Fig.~\ref{fig:AGN_UVLF}. We show the Poisson error on our combined LF as the orange shaded area.
We compare our LF to the $z \simeq 4$ observed LF of \citet{Adams2020}, and as expected from the \astraeus calibration, we have an excellent agreement with the galaxy UV LF at the faint end. We find that the \MUV regime where the galaxy and AGN UV LF overlap is around $\MUV \simeq -23$, similar to the observed LF. We note that while we have chosen to show the AGN UV LF including obscuration, the \citet{Merloni2014} obscuration fraction is below $f_{\rm obs} \lesssim 50\%$ for \MUV brighter than $-21.5$, so that the obscuration correction only plays a role in the \MUV range that is already dominated by the galaxy population.

\begin{figure}
  \centering
  \includegraphics[width=\columnwidth]{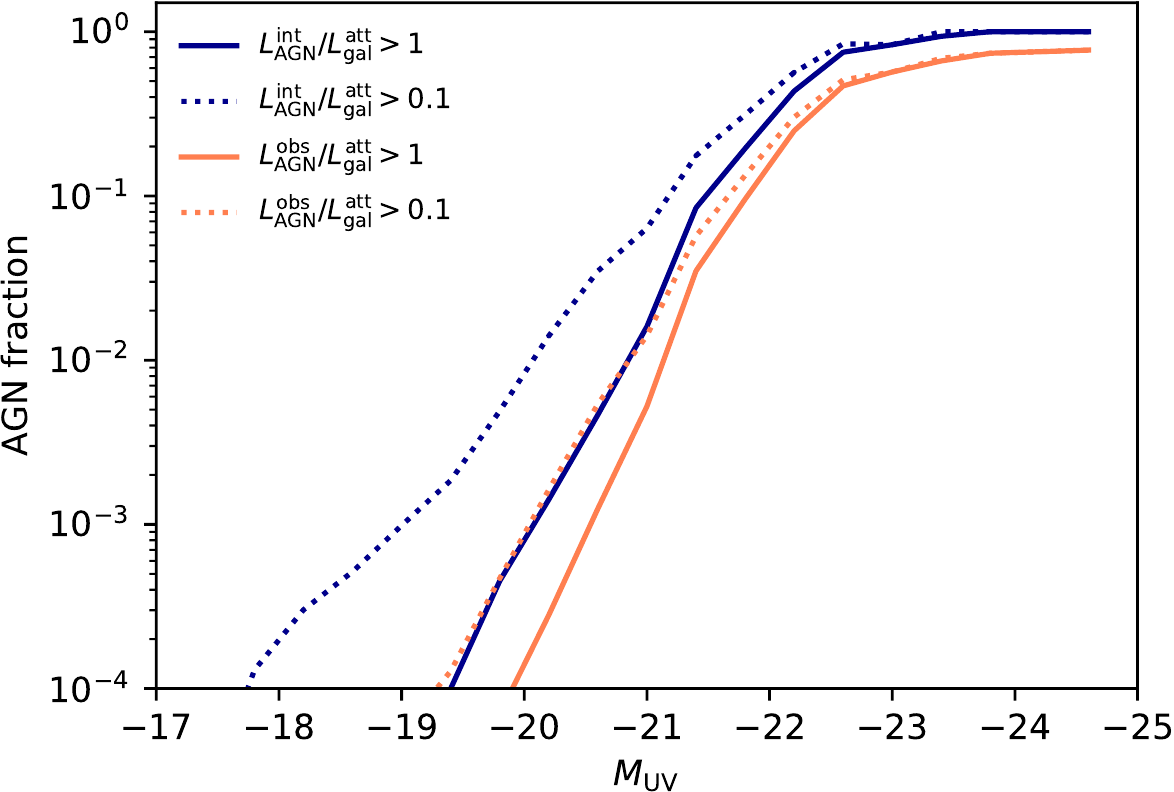}
  \caption{AGN fraction as a function of the attenuated galaxy \MUV at $z=4.5$, for different definitions of the AGN fraction: ratio of the intrinsic AGN UV luminosity to the attenuated galaxy UV luminosity in dark blue, including AGN obscuration in orange. The solid lines correspond to an AGN outshining the galaxy, while the dotted line correspond to an AGN with $10\%$ of the host UV luminosity.}
  \label{fig:Lgal_LAGN_UV}
\end{figure}
In that respect, our model is in good agreement with the empirical model of \citet{Volonteri2017} or with the numerical simulations of \citet{Trebitsch2020}, who found that at $z = 6$ the AGN dominated over the galaxy UV luminosity at \MUV brighter than $-23$. Interestingly, \citet{Sobral2018} found a similar critical \MUV at $z \simeq 2-3$, hinting at a slow evolution of this AGN-galaxy transition in the high-redshift Universe.
We can quantify this further by measuring the ratio of AGN to galaxy UV luminosity in the $f_{\rm seed} = 0.1, \fedd^{\rm high} = 1$ model, as shown in Fig.~\ref{fig:Lgal_LAGN_UV} as a function of the (attenuated) galaxy \MUV. The solid (dotted) lines mark the fraction of haloes where the AGN luminosity exceeds $100\%$ ($10\%$) of the attenuated galaxy UV luminosity, with the dark blue lines using the intrinsic AGN luminosity and the orange lines taking the obscured fraction into account.
Qualitatively, our results are consistent with the observations of \citet{Sobral2018} at $z \simeq 2-3$, who found that the AGN fraction $f_{\rm AGN} \gtrsim 50\%$ at $\MUV \simeq -21.5$. Quantitatively, our $f_{\rm AGN}$ is a bit lower than theirs for the $f_{\rm seed} = 0.1$ model, and assuming a higher $f_{\rm seed} = 1$ gives a critical \MUV much closer to the  \citet{Sobral2018} results.
Similar results have also been found by \citet{Piana2021b} using the parent \delphi model.
The model of \citet{Volonteri2017} yields a lower AGN fraction at high luminosity, mostly driven by the assumption that only $25\%$ of the galaxies are hosting active BHs. Nevertheless, they find that their AGN fraction saturates around $\MUV \simeq -22$, close to our findings.

\begin{figure*}
  \centering
  \includegraphics[width=\columnwidth]{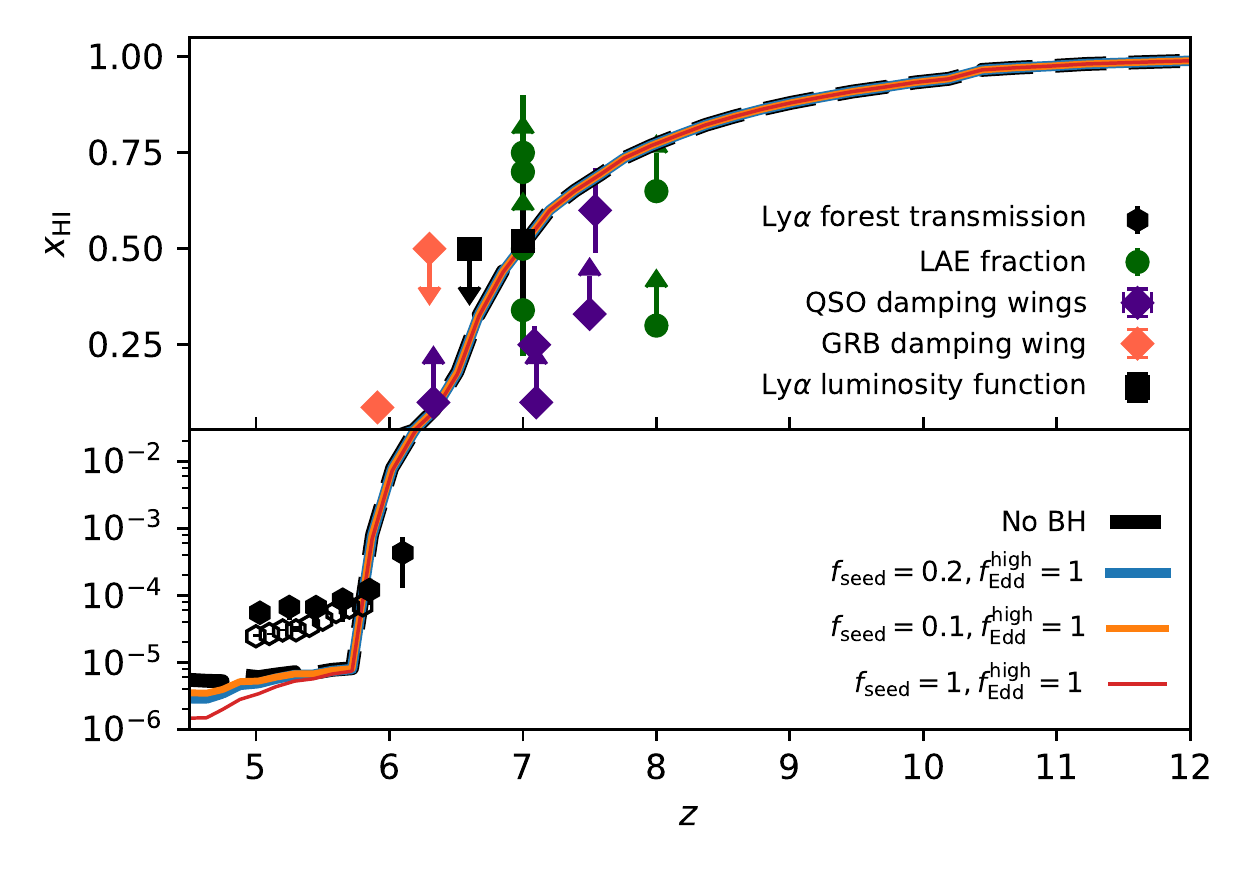}
  \includegraphics[width=\columnwidth]{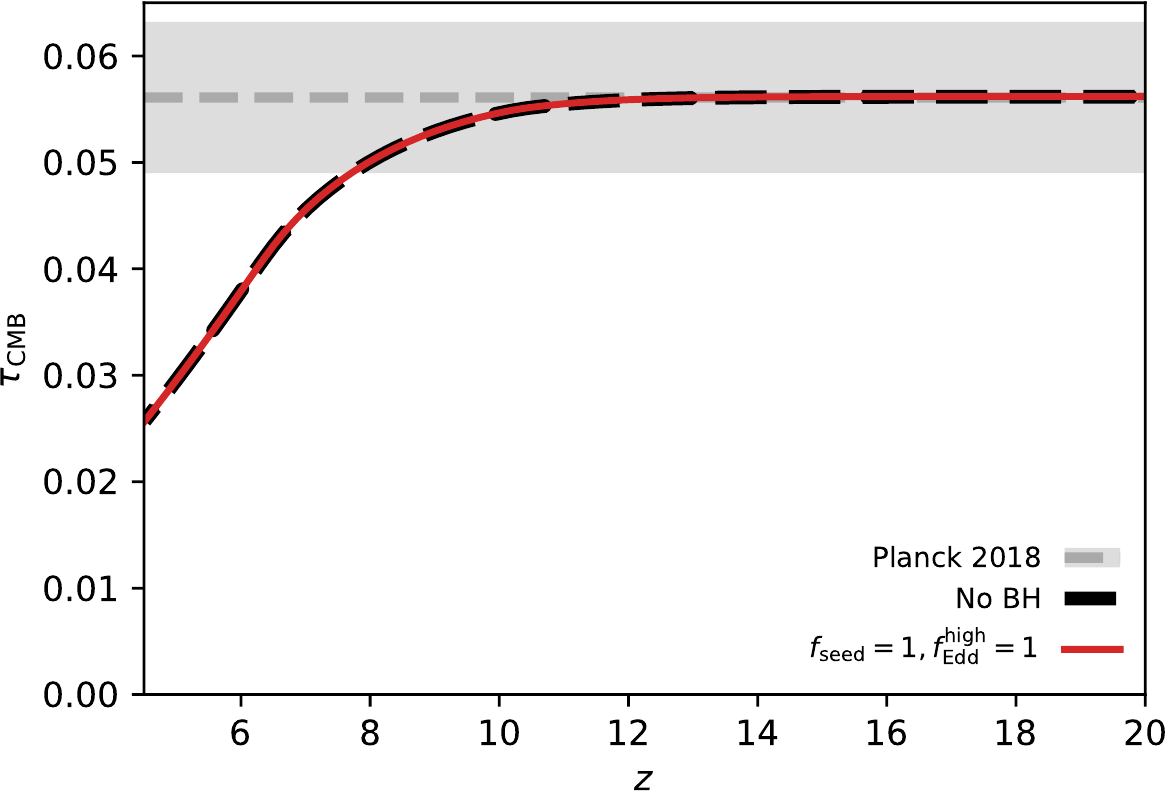}
  \caption{\emph{Left}: Evolution of the neutral fraction with redshift for our three main models (solid lines) compared to the baseline case without AGN (dashed black line). The points correspond to observational constraints on the neutral fraction (see text for details). All models assume $\fescAGN = 1$. \emph{Right}: CMB Thomson optical depth for our most extreme AGN case ($f_{\rm seed} = 1, \fedd^{\rm high} = 1, \fescAGN = 1$) compared to the case without AGN. The grey area indicates the 2018 \citet{Planck2018} constraints, used to calibrate the galaxy escape fraction model.}
  \label{fig:history}
\end{figure*}

\section{AGN contribution to reionization}
\label{sec:results}

Now that we have established the properties of our simulated AGN population, we turn our attention to their contribution to the reionization of the Universe in the \astraeus framework.

\subsection{Reionization history}
\label{sec:results:reionization}

We show in Fig.~\ref{fig:history} the reionization history resulting from our \astraeus simulations. The left panel focuses on the evolution of the neutral fraction $x_{\textsc{Hi}}$ for our three models with $\fedd^{\rm high} = 1$, all assuming $\fescAGN = 1$ as an extreme scenario, chosen to highlight the maximum effect of AGN on reionization allowed by our model.
The thick black dashed line corresponds to the original \astraeus model with no AGN contribution.
The other symbols mark observational constraints: black hexagons for measurements of the Lyman-$\alpha$ forest transmission from \citet[][full symbols]{Fan2006} and \citet[][open symbols]{Bosman2022}; green circles for constraints on the IGM opacity from the fraction of Lyman-$\alpha$ emitters in Lyman-break galaxy samples \citep[][]{Schenker2014, Ono2012, Pentericci2014, Robertson2013, Tilvi2014}; purple diamonds for measurements from quasar damping wings \citep{Mortlock2011, Schroeder2013, Banados2018, Durovcikova2020}; orange diamonds for gamma-ray bursts constraints \citep[][]{Totani2006, Totani2016}; and the black squares are constraints derived from the evolution of the Lyman-$\alpha$ luminosity function by \citet{Ouchi2010, Ota2008}. A fraction of these data points have been taken from compilation of \citet{Bouwens2015}.
Overall, at $z\gtrsim 6$, our models all match reasonably well the observational constraints, while we predict a too low residual neutral fraction in the post-reionization era.
The different AGN models show very little difference with the scenario without any AGN contribution, even for the most extreme $f_{\rm seed} = 1$ and $\fescAGN = 1$ model.
Despite significant differences in the modelling of the BH physics, our findings are remarkably consistent with the results of the \textsc{Dragons} project \citep{Qin2017}, who relies on the \textsc{Meraxes} semi-analytical model \citep{Mutch2016}. Similar to what we present here, they have found that the inclusion of an AGN component in their reionization model makes no difference to the evolution of their neutral fraction. The \textsc{Meraxes} model assumes a distinction between hot and cold gas, feeding the BH growth at different rates, while we assume that the BH growth proceeds at a rate that only depends on the halo mass and available global gas reservoir: the fact that our results are very similar suggests that the details of the BH modelling is largely irrelevant to estimate the AGN contribution to cosmic reionization.

We illustrate this further on the right panel of Fig.~\ref{fig:history}, which shows the Thomson optical depth from the CMB for our two most extreme cases compared to the confidence interval from \citet{Planck2018}. In practice, we measure the CMB optical depth following \citetalias{Hutter2021}:
\begin{equation}
  \label{eq:tau_cmb}
  \tau_{\rm CMB}(z) = \sigma_{\rm T} \int_0^z n_e(z') \frac{c}{(1+z')H(z')}\, \mathrm{d}z',
\end{equation}
where $H(z)$ is the Hubble parameter at $z$ and $n_e(z)$ is the electron number density at $z$, determined from the mass-weighted ionized fraction and the hydrogen and helium number densities. Since we do not track helium ionization, we assume that the fraction of singly ionized helium is the same as the hydrogen ionized fraction, and that helium is doubly ionized below $z<3$.
Even assuming the most optimistic BH seeding scenario, the AGN contribution remains negligible. This is predominantly because in our models, BH growth happens too late, so that the AGN contribution to the ionizing UV background only starts to be significant at $z \lesssim 5.8$ when reionization is mostly finished. We discuss this low-$z$ behaviour further in Appendix~\ref{app:lowz-uvb}.

\subsection{Source properties}
\label{sec:results:sources}

\subsubsection{Population-averaged properties}
\label{sec:results:sources:global}

\begin{figure}
  \centering
  \includegraphics[width=\columnwidth]{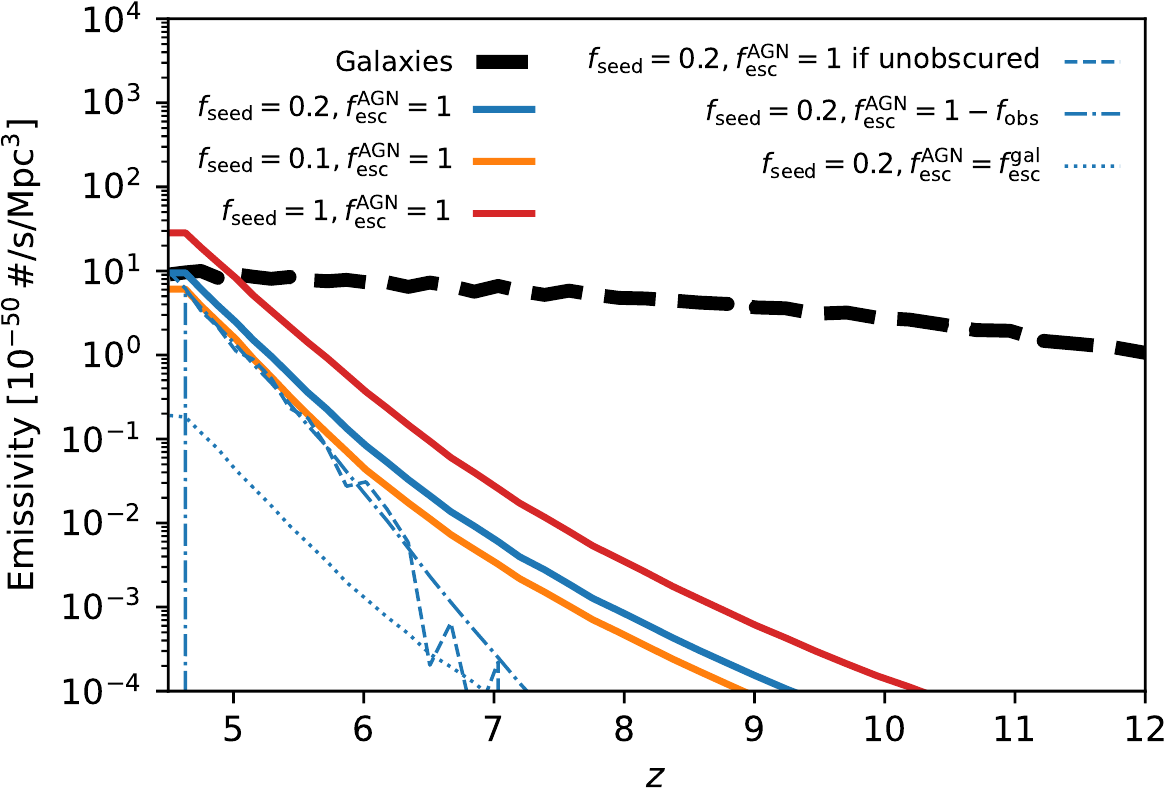}\vspace{-1em}
  \caption{Evolution of the escaping ionizing emissivity of galaxies (black dashed line) and AGN for the different scenarios considered in this work. The colours follow the same $f_{\rm seed}$ convention as in previous Fig.~\ref{fig:AGN_LbolLF}. Solid lines correspond to $\fescAGN = 1$, while the thin dashed line shows the model where $\fescAGN = 1$ for unobscured AGN, the thin dash-dotted line is $\fescAGN = 1-f_{\rm obs}$, and the thin dotted line assumes $\fescAGN = \fescstar$.}
  \label{fig:emissivity}
\end{figure}

Equipped with this understanding of how reionization proceeds in our model, we can now focus on the sources themselves. We show in Fig.~\ref{fig:emissivity} the ionizing emissivity that escapes into the IGM for galaxies (thick black dashed line) and for different AGN models. The colour still indicates $f_{\rm seed} = 0.1$ (orange), $f_{\rm seed} = 0.2$ (blue), and $f_{\rm seed} = 1$ (red), and we explore different \fescAGN models presented in Sect.~\ref{sec:methods:agn:radiation} with different line styles. The solid line assumes $\fescAGN = 1$, the dashed line is for the model where $\fescAGN = 1$ for unobscured AGN, the dash-dotted line shows $\fescAGN = 1 - f_{\rm obs}$, and the dotted line assumes $\fescAGN = \fescstar$. For clarity, we only show the variations of the \fescAGN model for the $f_{\rm seed} = 0.2$ case.

As expected from the discussion on $\Gamma_{\textsc{Hi}}$, the AGN emissivity stops being negligible only at $z \lesssim 6$, eventually taking over the galaxy contribution at the very end of our simulation. The higher normalisation of the runs with higher $f_{\rm seed}$ directly comes from the larger number of accreting BHs, which can be read from the bolometric LF in Fig.~\ref{fig:AGN_LbolLF}. The model with $\fescAGN = \fescstar$ yields a much lower emissivity, driven by the fact that the brighter AGN are predominantly hosted in high-mass galaxies, where SN feedback is less efficient, so that $\fescstar$ will be low in our model. For the two models relating \fescAGN to the \citet{Merloni2014} obscured fraction, the results are very similar, because the model where $\fescAGN = 1$ for unobscured AGN is essentially a random realisation of the model where $\fescAGN = 1 - f_{\rm obs}$. For these two models, the evolution of the AGN emissivity is steeper than for all other \fescAGN models. At $z \gtrsim 6-7$, we predict that the number density of bright AGN dramatically drops, so that most of the objects contribution to the AGN emissivity will be fainter, and therefore more obscured. At later times, the AGN emissivity is dominated by brighter, less obscured sources, so the global emissivity will resemble more the $\fescAGN = 1$ case.
Overall, our estimate of the ionizing emissivity from both galaxies and AGN are in good agreement with the model of \citet{Yung2021}, who found that the AGN contribution at $z \simeq 6$ was of the order $1-10\%$ depending on the assumed \fescAGN.
We find also a good consistency with the earlier results of \citet{Dayal2020} who found that the cumulative contribution of AGN reached 10\% to 25\% of the total emissivity depending on the assumed \fescAGN.
\begin{figure}
  \centering
  \includegraphics[width=\columnwidth]{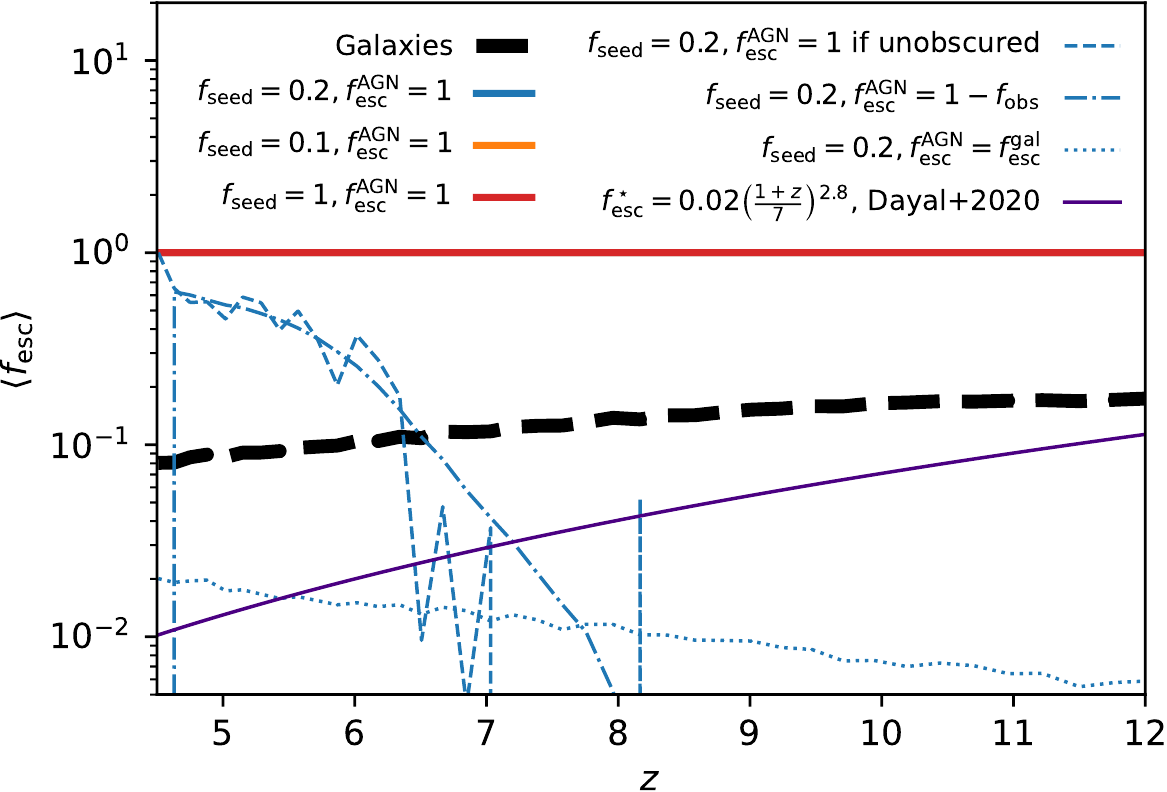}
  \caption{Population-averaged luminosity-weighted ionizing escape fraction for the galaxies (dashed black line) and the AGN with the same legend as in Fig.~\ref{fig:emissivity}. The solid purple line corresponds to the $\fescstar(z)$ model used in \citet{Dayal2020}.}
  \label{fig:fesc}
\end{figure}

Dividing the escaped emissivity by the intrinsic emissivity, we get the population-averaged escape fraction $\langle\fesc\rangle$, which we show in Fig.~\ref{fig:fesc} using the same legend as in Fig.~\ref{fig:emissivity}. We also added the fiducial \fescstar model from \citet{Dayal2020} as the thin violet line for comparison. Overall, $\langle\fescstar\rangle$ has a very mild evolution, with a slow decline with decreasing redshift. This is indicative that as cosmic time goes, more and more massive galaxies start to dominate the ionization budget. In contrast, the model with $\fescAGN = \fescstar$ evolves the other way: this would indicate that at lower $z$, the AGN that are contributing the most ionizing photons are located in lower mass galaxies, which can be understood easily since at fixed stellar mass, galaxies tend to host more and more massive BHs at lower $z$ (see Fig.~\ref{fig:MstarMBH}).
For the models linking \fescAGN to the obscuration fraction, we find again the behaviour from Fig.~\ref{fig:emissivity}: $\langle\fescAGN\rangle$ evolves from a very low value at high-$z$, when the AGN are mostly obscured, to a value of around $50\%$ at $z \lesssim 5$ when the dominant AGN contribution comes from brighter AGN, with a lower obscuration fraction.

\subsubsection{Which AGN contribute the most?}
\label{sec:results:sources:details}

\begin{figure}
  \centering
  \includegraphics[width=\columnwidth]{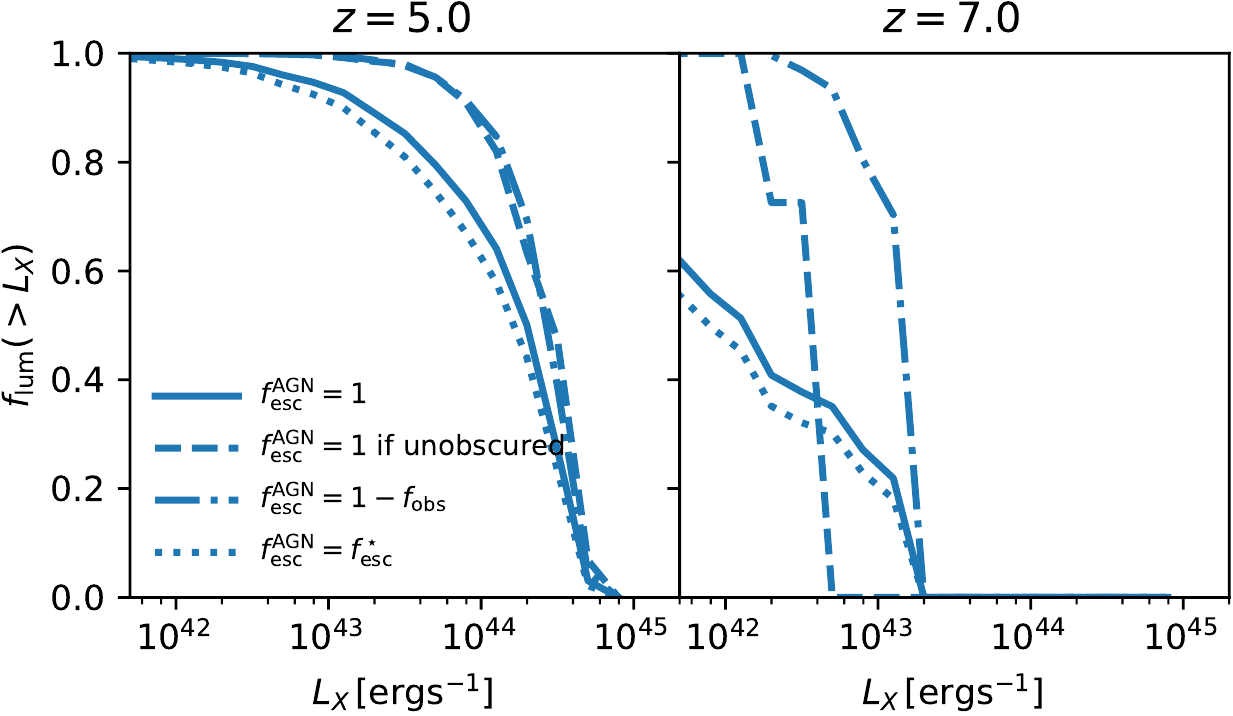}
  \caption{Fraction of the ionizing luminosity escaping from AGN brighter than a given X-ray luminosity $L_X$ at $z=5$ (left) and $z=7$ (right) for different \fescAGN models for the $f_{\rm seed} = 0.2$ scenario, using the same legend as in Fig.~\ref{fig:emissivity}.}
  \label{fig:EmAGN_contrib}
\end{figure}
We explore this more quantitatively in Fig.~\ref{fig:EmAGN_contrib}, where we measure the fraction of the total ionizing emissivity produced by AGN brighter than a given X-ray luminosity $L_X$ at $z=5$ (left) and $z=7$ (right), for the different \fescAGN models considered in this work. The scenarios where $\fescAGN = 1$ and $\fescAGN = \fescstar$ show very similar behaviours, with AGN fainter than $L_X \lesssim 10^{44}\,\mbox{erg}\,\mbox{s}^{-1}$ account for around $30\%$ of the ionizing luminosity at $z=5$. In contrast, the models relating \fescAGN to the obscuration fraction show a much more significant contribution from bright sources, with less than $10\%$ of the ionizing photons coming from AGN below that luminosity. This is a direct consequence of the shape of the obscuration fraction from \citet{Merloni2014}: we see from eq.~\ref{eq:fobs_merloni} that below $L_X \lesssim 10^{44}\,\mbox{erg}\,\mbox{s}^{-1}$, most AGN are (optically) obscured, while it is the case only for a small fraction of them above this luminosity.
At higher redshift, most of the AGN are below this critical luminosity and are therefore obscured: this explains why $\fescAGN \ll 1$ at $z \gtrsim 6$ for the models using $1 - f_{\rm obs}$ as a proxy for \fescAGN.

\begin{figure}
  \centering
  \includegraphics[width=\columnwidth]{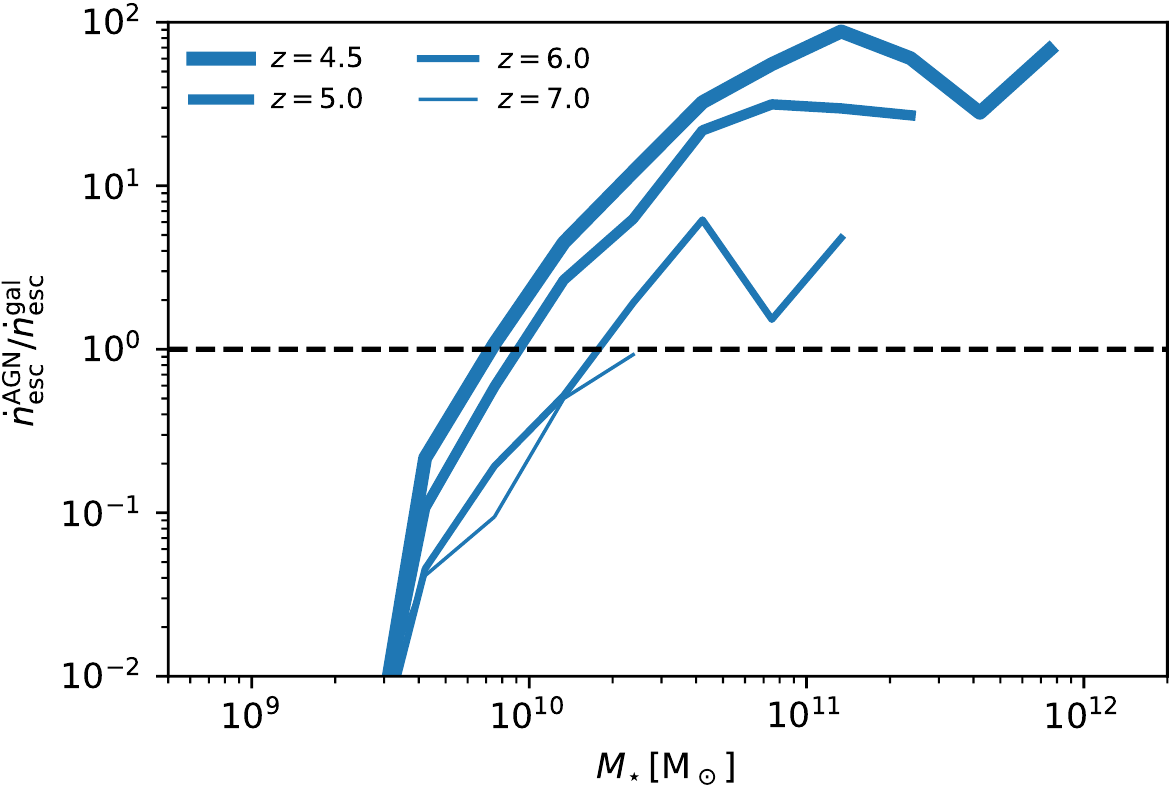}
  \caption{Ratio of the AGN to galaxy escaped emissivity as a function of stellar mass for the $f_{\rm seed} = 0.2, \fescAGN = 1$ at $z=4.5, 5, 6, 7$ indicated by lines of decreasing thickness. The AGN are contributing more ionizing photons in galaxies more massive than $\Mstar \simeq 10^{9.8 (10.3)}\,\Msun$ at $z=4.5$ ($6$).}
  \label{fig:ratio}
\end{figure}
Finally, we wish to find and characterise the regime in which the AGN radiation dominates over the stellar light. We quantify this in Fig.~\ref{fig:ratio} for the $f_{\rm seed} = 0.2, \fedd^{\rm high} = 1, \fescAGN = 1$ model, where we measure the ratio of the (escaped) ionizing emissivity from the AGN and its host galaxy as a function of the host stellar mass, at different redshifts (from $z=4.5$ to $z=7$, indicated with decreasing line thickness with increasing $z$). We find that the AGN starts to be the dominant source of ionizing photons in galaxies more massive than $\Mstar \simeq 10^{9.8}\,\Msun$ at $z=4.5$ ($\Mstar \simeq 10^{10.3}\,\Msun$ at $z=6$) , with the cut-off mass decreasing at lower redshift. This is qualitatively consistent with the results of \citet[][see their Figs.~2 and 5]{Dayal2020} at $z \gtrsim 6$, but about an order of magnitude higher than the cut-off mass of $\Mstar \simeq 10^9\,\Msun$ at lower redshift. We attribute this to a difference in the way we model the stellar component.
In \citet{Dayal2020}, $\fescstar$ is assumed to scale with redshift independently of the stellar mass of the galaxy. Compared to our implementation, this results in an average $\langle\fescstar\rangle$ significantly lower, with a steeper evolution with redshift.
Additionally, the fiducial model assumes a slightly lower maximum star-formation efficiency compared to us ($f_\star = 2\%$ compared to $2.5\%$), so that their galaxies will be slightly less massive than ours, on average.
Finally, their SN feedback is weaker than ours, with a coupling parameter $f_{w}^\star = 0.1$ vs $0.2$ for us. As in our model \citep[see e.g.][for a discussion on this \fescstar model]{Hutter2021b}, \fescstar saturates in galaxies where the SN feedback can eject all of the remaining gas, lowering value of $f_{w}^\star$ would result lowering the stellar mass threshold above which $\fescstar$ is decreasing, so that the (escaped) luminosity at fixed stellar mass would be lower. As a consequence, the relative contribution of the AGN would be higher.

\subsection{Reionization morphology}
\label{sec:results:morphology}

\begin{figure*}
  \centering
  \includegraphics[width=\linewidth]{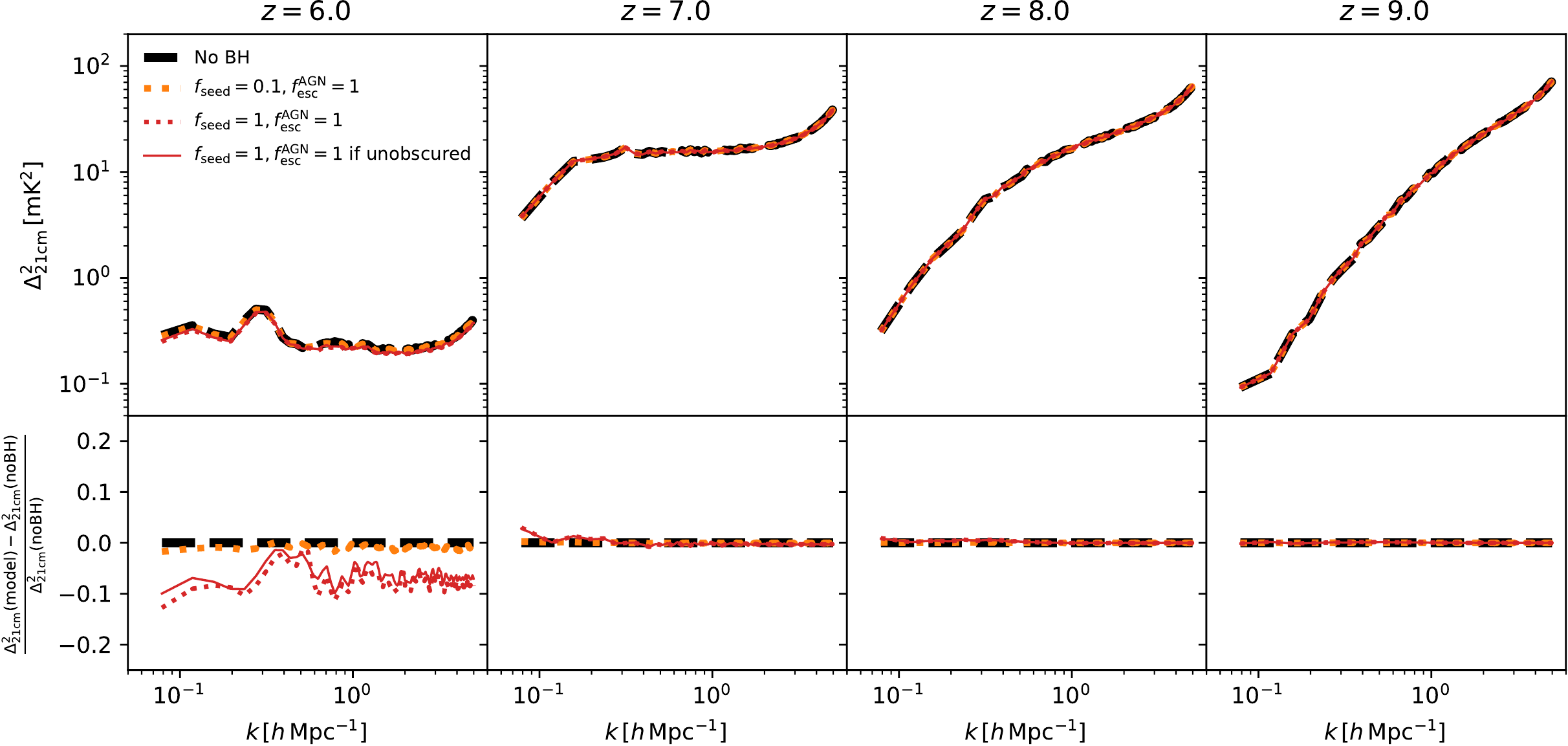}
  \caption{\emph{Top}: 21 cm power spectrum $\Delta^2_{21\rm cm}$ at $z$ from 6 to 9 (from left to right) for the models with no AGN (dashed black line), with $f_{\rm seed} = 0.1, \fescAGN = 1$ (dotted orange line), $f_{\rm seed} = 1, \fescAGN = 1$ (dotted red line), and $f_{\rm seed} = 0.1, \fescAGN = 1$ for unobscured AGN only (solid red line). \emph{Bottom}: fractional difference with the model without AGN. Overall, the AGN effect on the 21 cm signal is minor at best, and only at low redshift when the global signal is already very weak.}
  \label{fig:21cm}
\end{figure*}

Having established that statistically, AGN contribute very little to the overall photon budget of reionization, we now try to answer the question of whether the presence of rare but bright sources has any effect on the way reionization proceeds spatially. We approach this statistically, by measuring the power spectrum of the 21 cm signal following the approach of \citet{Hutter2020,Hutter2021}. Assuming that the spin temperature is well above the CMB temperature at the redshift of interest (a reasonable assumption at the later stages of reionization), the differential 21 cm temperature brightness $\delta T_b$ is given at any position $\mathbf{r}$ of the volume by
\begin{equation}
  \label{eq:21cm}
  \delta T_b(\mathbf{r}) = T_0 \left(1+\delta(\mathbf{r})\right) x_{\textsc{Hi}}(\mathbf{r})
\end{equation}
where
\begin{equation}
  \label{eq:T0_21cm}
  T_0 = 28.5\,\mbox{mK} \left(\frac{1+z}{10}\right)^{1/2} \frac{\Omega_b}{0.042} \frac{h}{0.73} \left(\frac{\Omega_m}{0.24}\right)^{-1/2}
\end{equation}
and $\delta(\mathbf{r})$ is the local gas over-density. From this, we compute and show in the upper panel of Fig.~\ref{fig:21cm} the 21 cm power spectrum $\Delta^2_{21\rm cm}$ at $z = 6, 7, 8, 9$ (from left to right). In each panel, the dashed black line corresponds to the model without any AGN contribution, the dotted orange line is the model with $f_{\rm seed} = 0.1, \fescAGN = 1$, the dotted red line shows $f_{\rm seed} = 1, \fescAGN = 1$, and the solid red line marks the scenario where $f_{\rm seed} = 0.1, \fescAGN = 1$ for unobscured AGN only. We view the last two scenarios to be the most extreme ones possible: the former maximises the AGN contribution, while the latter puts more emphasis on the spatial segregation of the sources.
As expected, the overall evolution of $\Delta^2_{21\rm cm}$ show a stronger signal on small scales at high $z$, before reionization is complete, and gets shallower on scales smaller than the ionized regions with increasing ionized fraction, until the full volume is reionized. At that stage, the signal becomes weak at all scales and depends on the residual neutral fraction.

We find that none of the AGN models, even the most extreme ones, have any significant impact on the 21 cm power spectrum.
The lower panel quantifies this as the fractional difference between the AGN models (orange and red) and the no AGN model. At all epochs considered, the AGN have virtually no impact on the power spectrum. The only (numerically) significant difference happens at $z = 6$, where the power spectrum is about $10\%$ weaker than without AGN. This is primarily driven by a slightly lower neutral fraction, but the absolute value of $\Delta^2_{21\rm cm}$ is very low at this epoch, causing this difference to be negligible in practice. While this seems in contradiction with the results of e.g. \citet{Kulkarni2017}, who find a strong imprint of AGN on the 21 cm power spectrum, the difference essentially comes from the very low contribution of AGN to reionization in our model.

This very limited effect of AGN on the 21 cm power spectrum may seem at odd with the picture in which rare, bright quasars are ionizing their immediate surrounding \citep[e.g.][]{Cen2000}, even imprinting a specific pattern on the 21 cm signal \citep{Bolgar2018}. However, we note here that because of the volume we survey in this work, we do not model the extremely bright but extremely rare quasars with bolometric luminosities exceeding $L_{\rm bol} \gtrsim 10^{47}\,\mbox{erg}\,\mbox{s}^{-1}$ powered by $\Mbh \gtrsim 10^{9}\,\Msun$ deep in the Epoch of Reionization, such as those found by \citet{Banados2018}, \citet{Yang2020} and \citet{Wang2021} at $z \gtrsim 7.5$. For these very early sources, which have very low number densities, we might expect a much stronger effect on the 21 cm morphology. That being said, the intrinsic scarcity of these extremely luminous quasars will not strongly affect our results on the global contribution of AGN to reionization, although it may impact the thermal and ionization state of the gas in their vicinity.

\section{Conclusions}
\label{sec:conclusions}

In this work, we have investigated the role of high-redshift AGN population in the reionization history of the Universe. For this purpose, we have implemented a model for the formation, growth, and feedback from SMBHs in the \astraeus framework, which allowed us to follow self-consistently the ionizing output of the evolving AGN population at $z \gtrsim 4.5$. We applied this framework to the \vsmdpl cosmological N-body simulation, which tracks the evolution of matter in a $(160 h^{-1})\,\mbox{Mpc}^3$ volume resolving haloes down to $M_{\mathrm{vir,min}} = 1.24\times 10^{8}h^{-1}\,\Msun$. {In addition to reproducing all key observable for galaxies at $z \geq 4.5$}, we have calibrated our AGN model to reproduce the observed bolometric LF at $z = 5$, and found that the resulting AGN population was in very good agreement with other high-redshift constraints, such as the BH mass function of \citet{Kelly2013} or various estimates of the AGN UV LF \citep{Glikman2011, Boutsia2018, Giallongo2019}. Moreover, the relative contribution of galaxies and AGN to the total UV luminosity of high-$z$ sources is well reproduced by our model as well.

Equipped with this robust model, we have been able to establish how AGN impact the establishment and maintenance of an ionizing background in the high-redshift Universe.
Our key findings are as follow:
\begin{enumerate}
\item The ionizing emissivity of AGN is too low to contribute significantly to the ionizing budget during the Epoch of Reionization, accounting only for 1-10\% of the escaping emissivity at $z=6$ depending on the assumed \fescAGN. This is mostly because the number density of AGN bright enough to produce a significant amount of ionizing photons is too low in the high-redshift Universe.
\item Taking into account the fact that a fraction of high-$z$ AGN are obscured further reduces the contribution of the overall AGN population to reionization, especially at the highest redshifts, when AGN are on average less luminous and more obscured.
\item AGN in the most massive galaxies ($\Mstar \gtrsim 10^{9.8-10.3}\,\Msun$ at $z=4.5-6$) can contribute more ionizing photons than their host, but this only comes into play significantly at $z \lesssim 6$, when reionization is complete.
\item Despite the fact that bright AGN do not have the same spatial distribution as the galaxies that predominantly reionize the Universe, we find virtually no impact of the AGN population on the global morphology of reionization, quantified by the 21 cm power spectrum.
\end{enumerate}
Overall this paint a picture in which AGN have an extremely limited impact on the reionization of the Universe, irrespective of the assumption we make on the escape of ionizing radiation from the AGN. This is in good agreement with numerous earlier works employing empirical models \citep[e.g.][]{Kulkarni2019}, semi-analytical models \citep[e.g.][]{Dayal2020}, or cosmological simulations \citep[e.g.][]{Trebitsch2021}. We stress that this is not in conflict with the studies of e.g. \citet{Grazian2018,Grazian2022} and \citet{Boutsia2021}: in our model, AGN do take over the UV background at $z \simeq 5$ (with variations around this value slightly depending on the assumptions for \fescAGN). 

We note that despite careful modelling, we found no DCBH seed in our set of simulations. We attribute this to two factors, predominantly. First of all, the \vsmdpl box is `only' $160 h^{-1}\,\mbox{Mpc}$ on a side: while this is enough to sample the properties of the galaxies during the Epoch of Reionization \citep[see e.g.][who estimated the importance of cosmic variance on reionization]{Ucci2021}, this is not quite enough to sample the extremely rare haloes, with a number density comparable to that of the brightest quasars ($\sim 10^{-9}\,\mbox{Mpc}^{-3}$). These sites are thought to be the birthplace of DCBH seeds, which would grow to become the most massive SMBH observed at $z \gtrsim 6$. This can be seen e.g. from our UV luminosity function, which stops around $\MUV \simeq -26$ at $z=4.5$, or even from the fact that we do not find any $\Mbh \simeq 10^{9}\,\Msun$ BH at $z = 6$ in our simulations. The second reason that may be causing the lack of DCBH seed in our simulation is the mass resolution of the \vsmdpl simulation. The minimum halo mass is $M_{\mathrm{vir,min}} = 1.24\times 10^{8}h^{-1}\,\Msun$, which is around the atomic cooling limit. While DCBH seeds are expected to form in haloes with masses of that order of magnitude, the formation history of these haloes is completely unresolved in our simulation. This is what led us to model the self-enrichment of these haloes following the approach of \citet{Trenti2007,Trenti2009}, but this is only a statistical approach. Taking into account a global LW background, we found that the probability for a starting halo to be pristine is of the order of $1\%$ at $z=15$, but this does not take into account the possibility for so-called `synchronised pairs' of haloes, where one halo would start forming stars earlier than its neighbour and its local LW flux would prevent star formation to occur there, therefore keeping this second halo prisitine and eligible for DCBH formation \citep[see e.g.][who explored the plausibility of such scenario]{Wise2019,Lupi2021}.

The first limitation could in principle be overcome by applying the \astraeus model on a large cosmological volume, such as the \textsc{Smdpl} and its $L_{\mathrm{box}} = 400 h^{-1}\,\mbox{Mpc}$ box. 
However, larger boxes come with the drawback that they typically have a lower mass resolution, worsening a lot that second issue. While we plan to investigate this in more details in a future work, we note that the inclusion of DCBH seeds from a larger simulation will certainly not change the overall results from this work. While this would likely result in the presence of several SMBH with masses in excess of $\Mbh \gtrsim 10^9\,\Msun$ at $z \gtrsim 6$, the quasars they would power would be too rare to significantly change the reionization history of the Universe. They might however leave some trace on the large-scale 21 cm power spectrum which will be observed with the SKA.

\section*{Acknowledgements}

We thank the anonymous referee for their useful comments and references which improved this manuscript.
MT, PD, SG and GY acknowledge support from the NWO grant 0.16.VIDI.189.162 (``ODIN''). AH, PD, LL, SG, and GY acknowledge support from the European Research Council's starting grant ERC StG-717001 (``DELPHI''). PD acknowledge support from University of Groningen's CO-FUND Rosalind Franklin Program. GY acknowledges financial support from MICIU/FEDER under project grant PGC2018-094975-C21.
We thank Peter Behroozi for creating and providing the \textsc{Rockstar} merger trees of the \vsmdpl. The authors wish to thank V. Springel for allowing us to use the \textsc{L-Gadget2} to run different Multidark simulations, including the \vsmdpl used in this work. The \vsmdpl simulation has been performed at LRZ Munich within the project pr87yi. The CosmoSim database (\url{https://www.cosmosim.org}) provides access to the simulation and the \textsc{Rockstar} data. The database is a service by the Leibniz Institute for Astrophysics Potsdam (AIP).
This work has made extensive use of the NASA's Astrophysics Data System, as well as the \textsc{Matplotlib} \citet{Hunter2007}, \textsc{Numpy/Scipy} \citet{Harris2020} and \textsc{IPython} \citet{Perez2007} packages.

\section*{Data Availability}

The results of the simulations presented in this work will be shared upon reasonable request to the corresponding author. The public version of the \astraeus code can be found in the \citet{Hutter2020code} ASCL entry. The \vsmdpl simulation upon which this work relies can be found at \url{https://www.cosmosim.org}.


\bibliographystyle{mnras}
\bibliography{astraeus} 



\appendix


\section{Photo-ionization rate and post-reionization neutral fraction}
\label{app:lowz-uvb}

\begin{figure}
  \centering
  \includegraphics[width=\columnwidth]{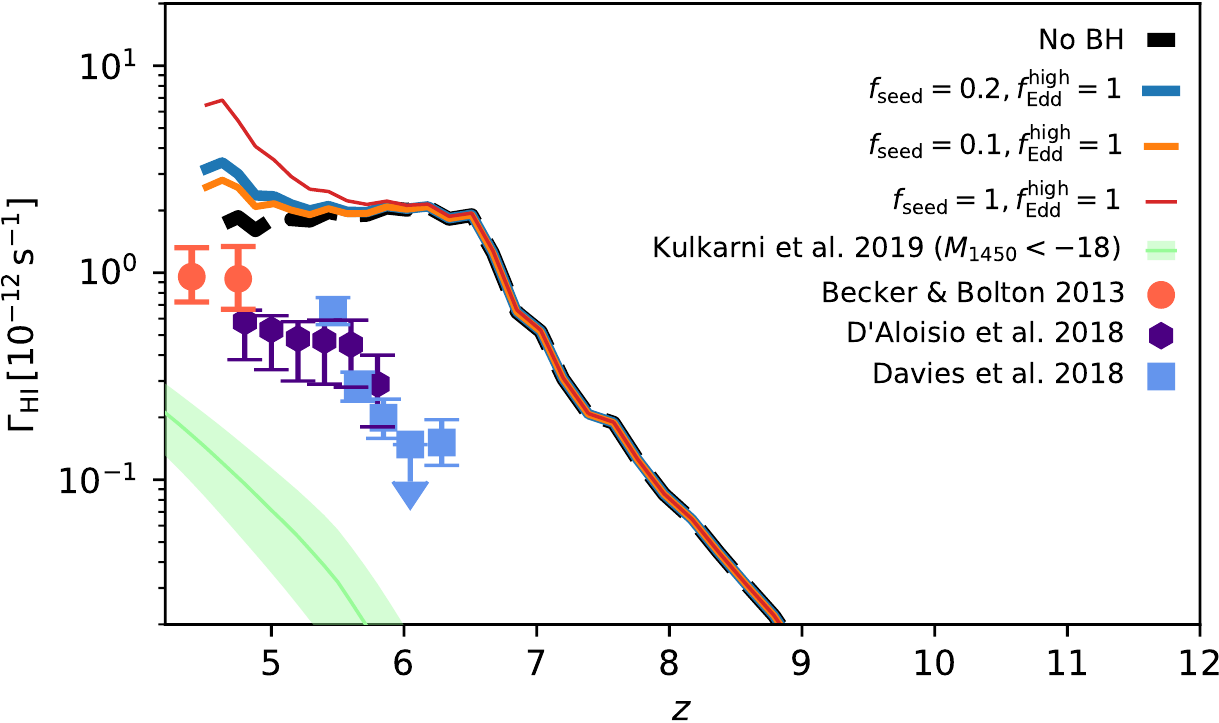}
  \caption{Evolution of the photo-ionization background $\Gamma_{\textsc{Hi}}$ for the same models as in Fig.~\ref{fig:history}. All models tend to over-estimate the $z \lesssim 6$ UV background, consistent with our under-estimation of the low-$z$ neutral fraction.}
  \label{fig:gamma}
\end{figure}

We show in Fig.~\ref{fig:gamma} the evolution of the \textsc{Hi} photo-ionization rate $\Gamma_{\textsc{Hi}}$ as a function of redshift for different models assuming $\fescAGN = 1$. The light green band indicates the AGN contribution to $\Gamma_{\textsc{Hi}}$ estimated from the \citet{Kulkarni2019} AGN UV LF. The red circles, purple hexagons, and blue squares indicate measurements of $\Gamma_{\textsc{Hi}}$ by \citet{Becker2013, DAloisio2018, Davies2018}, respectively.
We see that for all models that include AGN, $\Gamma_{\textsc{Hi}}$ only starts to significantly deviate from the scenario without AGN at $z \lesssim 5.5$, after reionization is complete. This is in very good agreement with the results of \citet{Trebitsch2021}, who estimate the contribution of AGN to the UV background in an overdense region of the Universe (therefore particularly favourable to BH growth), and find that their AGN population starts to dominate the UV background at $z \lesssim 4.5$. This is also consistent with the findings of \citet{Giallongo2019}, who find that at $z \simeq 5.6$ their observed AGN population could account for more than $20\%$ of the total UV background, and even higher at $z \simeq 4.5$.

At face value, however, the $\Gamma_{\textsc{Hi}}$ predicted from our simulation is significantly higher than the value inferred from observations, especially in the post-reionization era (where constraints exist). We interpret this as caused by the fact that we do not resolve small absorbers in our simulation. We use the `flux-based' method of \citet[][Sect.~2.2.2]{Hutter2018} to estimate the photo-ionization rate: at a distance $r$ from a single source, we have $\Gamma_{\textsc{Hi}}(r) \propto \dot{N}_{\rm ion} \exp(-r/\lambda_{\rm mfp}) / r^2$, where $\lambda_{\rm mfp}$ is the mean free path. Post-reionization, $\lambda_{\rm mfp}$ scales as $f_{\rm self-shielded}^{-2/3}$, where $f_{\rm self-shielded}$ is the volume fraction of self-shielded gas. Missing the dense, self-shielded absorbers in our simulations leads to an over-estimation of the post-reionization $\lambda_{\rm mfp}$, and therefore to an over-estimation of the photo-ionization rate.
This is directly related to the low post-reionization neutral fraction we see in Fig.~\ref{fig:history}: since we are missing the dense clumps that stay neutral even after reionization is complete, we end up over-estimating the photo-ionization rate. This could be in principle corrected by re-calibrating our mean free path model, but doing so would be resolution-dependent, and will not impact our results, so this is beyond the scope of this work.


\bsp	
\label{lastpage}
\end{document}